\documentclass[twocolumn,secnumarabic,amssymb, nobibnotes, aps, pra, 10pt]{revtex4-1}

\def\pra{{Phys.~Rev.~A}}

\def\prl{{Phys.~Rev.~Lett.}}

\def\nat{{Nature}}

\def\physscr{{Phys.~Scr}}

\usepackage{isomath}

\usepackage{upgreek} 

\usepackage{epsfig, amsmath,amsfonts, amssymb,graphicx,amsthm,color}

\usepackage{mathtools}
\usepackage{dcolumn}
\usepackage{bm}
\usepackage{rotating}
\usepackage{amssymb}
\usepackage[latin1]{inputenc}
\usepackage{comment}
\usepackage{multirow}
\usepackage{tabularx}
\usepackage{color}
\usepackage{longtable}

\newcommand {\Ket}[1]         {\ensuremath{| \, #1 \rangle}}

\begin{document}

\title{Laser and microwave spectroscopy of even-parity Rydberg states of neutral ytterbium\\and Multichannel Quantum Defect Theory analysis}
\author{H. Lehec, A. Zuliani, W. Maineult, E. Luc-Koenig, P. Pillet and P. Cheinet}
\email[Contact: ]{patrick.cheinet@u-psud.fr}
\affiliation{Laboratoire Aim\'{e} Cotton, CNRS, Univ. Paris-Sud, ENS Cachan, Universit\'e Paris-Saclay, B\^{a}t. 505, 91405 Orsay, France }
\author{F. Niyaz and T. F. Gallagher}
\email[Contact: ]{tfg@virginia.edu}
\affiliation{Department of Physics, University of Virginia, Charlottesville, Virginia 22904 }
\date{\today}

\begin{abstract}
\noindent\textbf{Abstract.} New measurements of high-lying even parity $6sns\, {}^1 \! S_0$ and $6snd\,{}^{3,1}\!D_2$ levels of neutral $^{174}$Yb are presented in this paper. Spectroscopy is performed by a two-step laser excitation from the ground state $4f^{14}6s^2 \, {}^1 \! S_0$, and the Rydberg levels are detected by using the field ionization method. Additional two-photon microwave spectroscopy is used to improve the relative energy accuracy where possible. The spectroscopic measurements are complemented by a multichannel quantum defect theory (MQDT) analysis for the J=0 and the two-coupled J=2 even parity series. We compare our results with the previous analysis of Aymar {\it{et al}} \cite{Aymar_1980} and analyze the observed differences. From the new MQDT models, a revised value for the first ionization limit $I_{6s}=50443.07041(25)$ cm$^{-1}$ is proposed.
\end{abstract}

\maketitle
\section{Introduction}
\label{sec:intro}


The Rydberg levels of Yb are a focus of attention for two reasons. First, two (valence) electron atoms are attractive for optical clocks based on the neutral atoms \cite{Ovsiannikov_2011, Bowden_2017} and their ions \cite{Safronova_2012, Tamm_2016}. The largest systematic frequency uncertainty in an optical ion clock arises from the black body radiation shift \cite{Jiang_2009}, which is determined by the ionic polarizabilities. These polarizabilities can be determined from accurate measurements of the quantum defects of the high $\ell$ $6sn\ell$ bound Rydberg levels converging to the ionic states of the clock transition \cite{Ward_1996}. Thus the accurate energy measurement of high $\ell$ Rydberg levels of Yb promises reduced uncertainties on the Yb$^+$ clock frequency.

The second attraction of the Rydberg states of Yb is related to cold Rydberg atom experiments. While ultra-cold atoms in their ground state have van der Waals interactions with a range of a few nanometers, due to their exaggerated properties ultracold Rydberg atoms have micrometer range interactions. The use of cold Rydberg atoms thus presents new possibilities in quantum physics. These include quantum simulation \cite{Weimer_2010} or quantum engineering \cite{Saffman_2010} with, for example, the realization of quantum gates using the dipole blockade \cite{Lukin_2001, Urban_2009, Gaetan_2009} or the production of single photons \cite{Dudin_2012}. The attraction for ytterbium is that once one electron is excited to a Rydberg state, the second valence electron, which is the single valence electron of the ionic core, is easily excited by a laser \cite{Bell_1991}. This excitation produces a doubly excited state which can decay radiatively or by auto-ionization. This experimental technique, called \emph{Isolated Core Excitation} (ICE) \cite{Cooke_1978}, offers several fascinating possibilities. One example is the optical imaging which could be performed by collecting fluorescence photons from the core. This would be similar to the imaging in \cite{McQuillen_2013, Lochead_2013} but without the need to ionize the Rydberg cloud. On one-electron Rydberg atoms, other imaging \cite{Schauss_2012, Gunter_2012} or trapping \cite{Dutta_2000, Anderson_2011, Saffman_2005} techniques have been implemented, relying on complex schemes or compromises.

To realize the full potential of ICE as a diagnostic and manipulation technique and to determine the core polarizabilities requires a comprehensive knowledge of the spectroscopy of the atom in question. Yb has been the subject of several optical studies with pulsed lasers, in particular on even parity levels \cite{Aymar_1980, Wyart_1979, Camus_1980, Xu_1994}, which have provided a wide range of spectroscopic data as well as multichannel quantum defect theory (MQDT) analysis of these series, but with a typical uncertainty on the GHz scale. In addition, similar measurements \cite{Aymar_1984} provide the currently accepted value for the first ionization limit $I_{6s}=50443.08(5)$ cm$^{-1}$. Extensive microwave measurements connecting the Yb $6sns$, $6snp$, and $6snd$ Rydberg series have also been reported for $40\leq n\leq 58$ \cite {Maeda_1992}. While these measurements are useful for $n\geq40$, they cannot be extrapolated towards lower $n$, where doubly excited states perturb the regularity of the bound $6sn\ell$ series, with $n$ the principal quantum number and $\ell$ the orbital angular momentum quantum number.

Here we present new spectroscopic measurements of high-lying even parity $6sns \, {}^1\!S_0$ and $6snd \, {}^{1,3}\!D_2$ Rydberg series of Yb. We have obtained an improved accuracy of two to three orders of magnitude on the absolute level energies and extended the microwave spectroscopy to significantly lower $n$. These levels can serve in the future as a reference for microwave spectroscopy of high $\ell$ Rydberg states. In the following sections we describe the experimental approaches and analysis methods, present our spectroscopic results, and describe the related MQDT analysis.

\section{Experimental Approach}
\label{sec:exp_setup}

\begin{figure}[ht]
\mbox{
\centering
\includegraphics*[width=0.95\columnwidth]{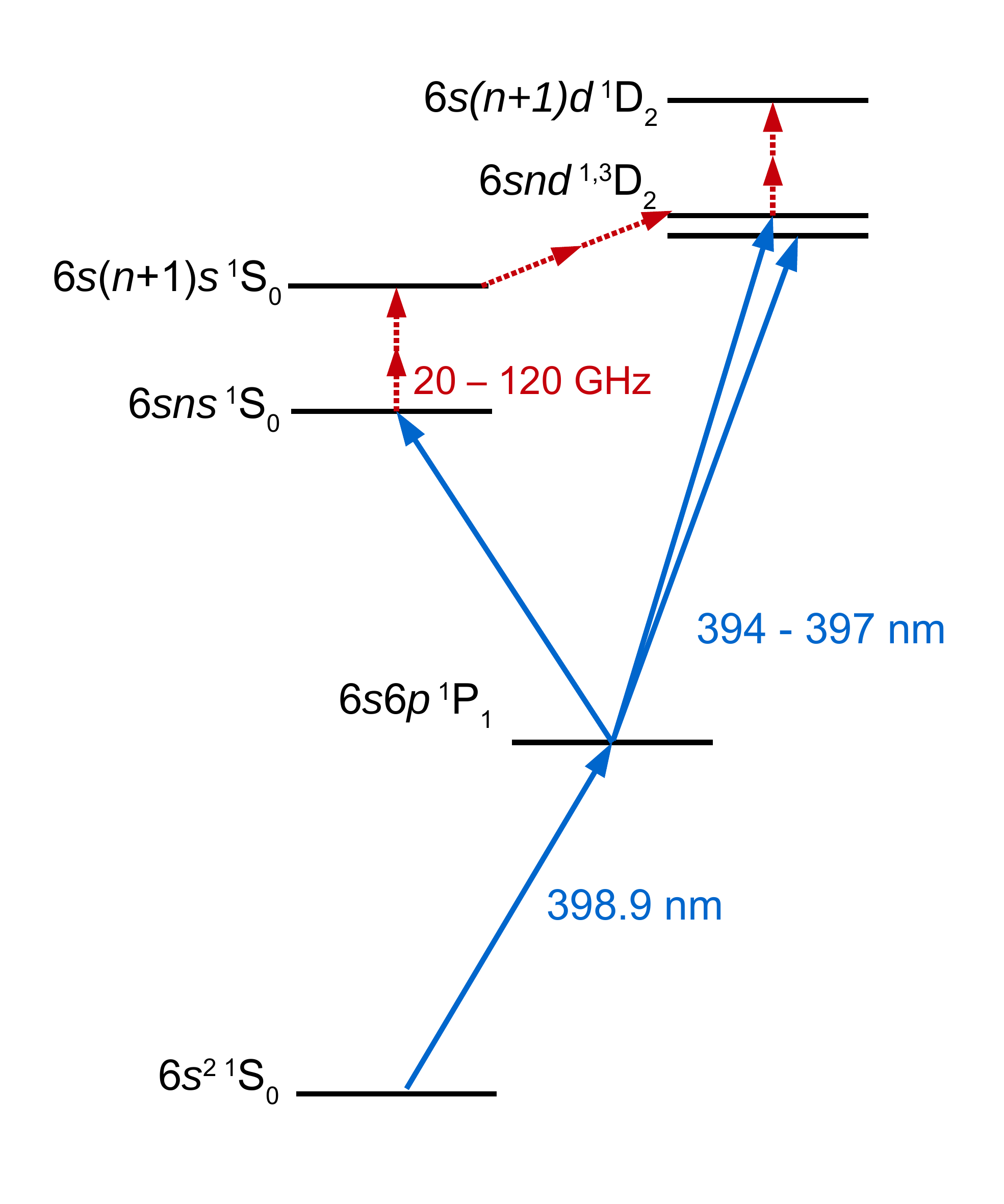} 
	}
\caption{Experimental energy level scheme. From the $6s^2\,{}^1\!S_0$ ground state, a first laser at around 398.9 nm excites atoms to the $6s6p\,{}^1\!P_1$ level. Then a second laser tunable around 396 nm excites the atoms to $6sns\,{}^1\!S_0$, $6snd\,{}^1\!D_2$ at UVA and also to $6snd\,{}^3\!D_2$ at LAC. For the singlet states, two-photon microwave transitions are performed at UVA, symbolized with red broken arrows. The blue arrows correspond to the optical spectroscopy measurements made at LAC.}
\label{fig_Excitation}
\end{figure}

We have performed joint spectroscopic measurements on two complementary setups. Both sets of measurements can be understood with the aid of Fig. \ref{fig_Excitation}. One cold atom setup in Laboratoire Aime Cotton (LAC) allows performing accurate optical spectroscopic measurements in order to obtain the absolute energies of the different observed Rydberg levels. The other experiment, at the University of Virginia (UVA), is performed on an atomic beam and allows measuring multi-photon microwave transitions between different Rydberg levels to determine their relative energy difference with better accuracy. We chose to perform our measurements on $^{174}$Yb, the most abundant isotope, as the quantum defects of high $\ell$ levels do not depend on the isotope. Both experiments use a two-photon excitation: from $6s^2\,{}^1\!S_0$ to $6s6p \,{}^1\!P_1$ level with a first laser at 398.9 nm and a second laser at around 396 nm to reach $6sns$ and $6snd$. The UVA experiment reaches $6sns\,{}^1\!S_0$ and $6snd\,{}^1\!D_2$ levels while LAC experiment also detects $6snd\,{}^3\!D_2$ levels. We now present the specific details of each experiment.

\begin{center}
\textbf{Microwave measurements}
\end{center}

In the microwave experiments at UVA, atoms in a thermal beam of natural Yb pass between two horizontal plates 1.5 cm apart, where they are excited to Rydberg levels by two copropagating laser beams. These lasers are pulsed tunable dye lasers running at a 20 Hz repetition rate. The laser beams cross the atomic beam at a right angle, defining the region in which the Yb Rydberg atoms interact with the microwaves. A $1\mu$s long microwave pulse, starting 50ns after the laser pulses, drives one of the transitions shown by the broken arrows of Fig. \ref{fig_Excitation}. The microwave field is generated by an Agilent 83620A synthesized sweep generator which produces a continuous wave output from 10MHz to 20GHz. A General Microwave DM862D switch is then used to produce the microwave pulses. Several frequency multipliers: a Narda DBS2640X220 active doubler, a Narda DBS4060X410 active quadrupler, a Pacific Millimeter V2WO passive doubler and a Pacific Millimeter W3WO or D3WO passive tripler were used to multiply the synthesizer frequency to the desired frequency. The microwaves propagate from the frequency multiplier through a WR28 waveguide feedthrough to a WR28 horn inside the vacuum chamber.

Approximately 50ns after the end of the microwave pulse a large negative voltage pulse is applied to the bottom plate to field ionize the Rydberg atoms and eject the freed electrons from the interaction region. The amplitude of the field pulse is chosen to allow temporal separation of the ionization signals from the initial and final states of the microwave transition. The freed electrons pass through a hole in the top plate and are detected by a Micro-Channel Plate (MCP) detector. The signal from the final state of the transition is recorded by a gated integrator as the microwave frequency is slowly swept across a resonance over many shots of the laser. The data are stored in a computer for later analysis. We have investigated three different two-photon microwave resonances: $6sns\,^1\!S_0\rightarrow6s(n+1)s\,^1\!S_0$, $6snd\,^1\!D_2\rightarrow6s(n+1)d\,^1\!D_2$, $6s(n+1)s\,^1\!S_0\rightarrow6snd\,^1\!D_2$.

The accuracy of the two photon microwave transitions can be limited by different error sources. The AC Stark shifts of the two photon transitions are generally small, typically 200 kHz for the maximum power. This effect is minimized measuring the resonances at several microwave powers to extrapolate the zero field frequencies with an uncertainty around 20 kHz. The frequency uncertainty of the synthesizer is less than 5 kHz, which leads to an uncertainty after frequency multiplication of less than 30 kHz in the worst multiplication case. DC Stark shifts can be important for Rydberg states.  For each initial state, we apply a voltage on the top plate to minimize this shift with an estimated uncertainty of around 100 kHz. The Fourier transform limited resonance width of 160 kHz allows to determine the resonance center with an uncertainty of around 20 kHz. The earth's magnetic field has a negligible effect on singlet-singlet transitions, as all $\Delta m=0$ transitions are unaffected by the magnetic field \cite{Shuman_2007}. Finally, the statistical noise on each measurement is around 60 kHz. We thus estimate the total uncertainty on our microwave measurements to be less than 200 kHz on the microwave frequencies and thus less than 400 kHz on the energy intervals. The uncertainty in \cite{Maeda_1992} is found to be around 250 kHz on the microwave frequencies, justifying the inclusion of these data in our analysis. 

\begin{center}
\textbf{Optical Measurements}
\end{center}

In the optical experiments at LAC, we have built an Yb cold atom experimental set-up similar to that used by Kuwamoto et al. \cite{Kuwamoto_1999} : a thermal beam of Yb, formed by heating a dispenser, passes through a Zeeman slower, where it is decelerated by light at the 398.9 nm $6s6p\,^1\!S_0\rightarrow6s6p\,^1\!P_1$ transition. A significant difference from the approach of Kuwamoto et al. is the introduction of a 2D molasses to counteract the beam divergence introduced by the Zeeman slower. This 2D molasses uses 556 nm light at the $6s^2\,^1\!S_0\rightarrow 6s6p\,^3\!P_1$ intercombination transition frequency. After the 2D molasses, the atoms are captured in a 3D magneto optical trap (MOT) using the same intercombination transition.

To excite the atoms to Rydberg levels, the first photon is provided by the Zeeman slower laser: a fraction of continuous laser light at 398.9 nm is diverted and sent through an acousto-optical modulator (AOM) to form pulses of around 500 ns duration at a 10 Hz repetition rate. They provide population in the $6s6p\,{}^1\!P_1$ state. The second photon, which couples the $6s6p\,{}^1\!P_1$ state to the desired Rydberg state, is provided by a frequency doubled Ti:Sapphire laser. Large $n$ Rydberg levels are detected by pulsed field ionization. A set of high-voltage electrodes provides a maximum electric field of around 600 V/cm in the Rydberg excitation volume, allowing the ionization of Rydberg states of $n\geq40$. For lower states, we use ICE with a laser at 369.5 nm on the $6s_{1/2}\rightarrow 6p_{1/2}$ ionic core transition \cite{Bell_1991} to convert bound $6sn\ell$ Rydberg atoms into autoionizing $6p_{1/2}n\ell$ atoms \cite{Cooke_1979}. In all cases, the resulting ions are detected with an MCP, and the time resolved signal is captured by a gated integrator as the frequency of the second laser is slowly swept over many shots of the laser.

We infer the Rydberg level energy from the sum of the $6s6p\,{}^1\!P_1$ intermediate level energy and the second photon energy. The energy of the intermediate level is taken to be 751 526 533.5 MHz according to the recent measurement in \cite{Kleinert_2016} with an uncertainty of 0.33 MHz. The corresponding laser in our experiment is servo-locked to the saturated absorption signal obtained on a separate spectroscopy vacuum cell where the Yb atom beam is saturated with the intercombination laser, providing a long term stability better than the transition linewidth of 200 kHz. The short term stability of this laser is limited to around 2 MHz due to the width of the saturated spectroscopic signal which is broadened to the $6s6p\,{}^1\!P_1$ linewidth of 30 MHz. The energy of the second photon is measured using a commercial wavemeter, \emph{High Finesse WS-Ultimate 10}, which we calibrate using the well known $D_2$ line of cesium at 852 nm available in our laboratory.  The wavemeter has a guaranteed accuracy of 10 MHz at three standard deviations within $\pm$ 200 nm around the calibration wavelength. To be within this range, we measure the Ti:Sa laser frequency before the doubling stage, at around 792 nm. The uncertainty after doubling is thus 6.7 MHz at one standard deviation. The repeatability of the measurement is announced to be around 2 MHz at one standard deviation. This leads, after frequency doubling, to an expected statistical noise of 4 MHz. Moreover, each measurement is rounded to 1 MHz before doubling. The two photon Rydberg excitation observed linewidth is 15 MHz, partially broadened by the intermediate state linewidth of 30 MHz, leading to an uncertainty in the line center of less than 1 MHz. The DC Stark shift is minimized using compensation electrodes to cancel the residual field. This field is canceled with the highest principal quantum number used, $n=80$, which is the most sensitive. The residual uncertainty is then estimated to be at most 200 kHz. Finally the combination of the MOT magnetic field gradient, which is not turned off during the measurement, and the effect of the Zeeman slower laser pushing the MOT away from the magnetic field center leads to a residual field of around 1 Gauss. But thanks to a linear polarization of our Rydberg excitation laser beams, the transitions shifted up and down should have equal weight leading to a broadening but no residual shift. Therefore we assume a magnetic field induced uncertainty of less than 100 kHz. The total uncertainty on the Rydberg energy is thus 7.5 MHz and the statistical noise is expected to be 5 MHz.

\section{Analysis Methods}
\label{sec:Analysis_Methods}

Rydberg spectra of alkali atoms, possessing a single Rydberg electron outside a spherically symmetric ionic core, are usually characterized with single channel quantum defect theory (QDT). It assumes that the electron moves in a Coulomb potential during most of its orbit except for short times when it comes close to the ionic core and collides with it. A channel is defined by fixing the quantum numbers of the ionic core, here singly charged, the orbital quantum number $\ell$ of the Rydberg electron and the coupling scheme defining the total angular momentum $J$. At any arbitrary energy $E$, the channel wave function is solution of the Schr\"odinger equation. Outside the ionic core ($r>r_c$) where the potential becomes purely Coulombic, the wave function is described as a superposition of the analytically known regular and irregular Coulomb functions $f$ and $g$:
$\cos[\pi\,\delta(E,\ell)]\,f(E,\ell,r)-\sin[\pi\,\delta(E,\ell)]\,g(E,\ell,r)$ where the quantum defect $\delta(E,\ell)$ is slowly varying with $E$. The corresponding phase $\pi \delta(E,\ell)$ represents the wave function phase shift with respect to the pure hydrogenic regular one $f(E,\ell,r)$ and is due to all short-range non-Coulombic interactions present at $r<r_c$. For a bound level of energy $E$ smaller than the first ionization limit $I$, one has to apply as boundary condition that the wave function vanishes at $r\rightarrow\infty$ which leads to the following equation for the energy:

\begin{equation}
E(n,\ell) = I - \frac{R}{\nu(n,\ell)^2}
\label{eq:Energy_eq}
\end{equation}

\noindent where R is the finite-mass corrected Rydberg constant and $\nu(n,\ell)=n-\delta (n,\ell)$ is called the effective quantum number. Within a specific $n\ell$ Rydberg series, the quantum defect varies slowly with the energy and is usually described with the Ritz formula defining the quantum defect $\delta_0 (\ell)$ at the ionization limit and additional higher order terms $\delta_i (\ell)$ accounting for its energy dependence in a Taylor series. For ease of use, it is expressed as a function of the principal quantum number and thus reads:

\begin{equation}
\delta (n,\ell) = \delta_0 (\ell) + \frac{\delta_1 (\ell)}{(n-\delta_0 (\ell))^2} + \frac{\delta_2 (\ell)}{(n-\delta_0 (\ell))^4} + ...
\label{eq:Ritz_eq}
\end{equation}

In two valence electron atoms such as alkaline-earth atoms, the Rydberg series $n_0s$ $n\ell$ converging toward the first ionization limit $n_0s$ can be perturbed by doubly-excited levels $n_0\prime\ell\prime$ $n_0\prime\prime\ell\prime\prime$ converging towards excited ionization limits $n_0\prime\ell\prime$, resulting in irregular variations of their quantum defects $\delta(n,\ell)$. Indeed, only the total angular momentum $J$ and parity $\Pi$ are exact quantum numbers, and exchange of angular momentum and energy between the two colliding electrons restricts the validity of the single channel QDT approach.

Multi-channel Quantum Defect Theory (MQDT) \cite{Fano_1975}, a generalization of Seaton's quantum defect theory \cite{Seaton_1966}, has been developed to accurately describe two electron atom spectra where a single electron $n\ell$ is excited to a Rydberg level, the second electron $n_0\ell_0$ with a low excitation being named the valence electron. MQDT provides an exact parametrization of the energy spectrum and of the wave functions outside the singly charged core $n_0\ell_0$ with radius $r_0$ in terms of channel coupling between $N$ channels by introducing  a small number of nearly energy-independent parameters with a physical meaning. This theory has been described in detail in many papers \cite{Fano_1975, Cooke_1985, Aymar_1996, Vaillant_2014}, therefore we do not present the complete mathematical framework but only the necessary concepts to understand our comparison with previous work \cite{Aymar_1980}. Our analysis is based on the 'eigenchannel' MQDT formulation \cite{Fano_1975} which introduces two sets of channels defined in two different ranges of the Rydberg electron distance from the ionic core and related by a unitary transformation matrix $U_{i\alpha}$: The $N$ 'collision channels', labelled $i$, describing at large distance the Coulombic two-body interaction between the singly charged ionic core and the outer electron and the $N$ 'eigenchannels', labelled $\alpha$, suitable to describe at short-distance the three-body interactions between the doubly charged ionic core of radius $r_c$ and the two electrons. These interactions are responsible for the inelastic scattering of the Rydberg electron between the different ionization channels.

More explicitly, a particular collision channel $i$, defined at a distance $r>r_0$ where the interaction between the Rydberg electron and the singly charged ion becomes purely Coulombic, is similar to the single channel introduced in QDT. It describes an incident electron with orbital angular momentum  $\ell_i$ colliding with a specific level of the ionic core with energy $I_i$ and total angular momentum $J_{c_i}$. The specification of the intermediate quantum numbers defining the coupling scheme used to construct the total angular momentum $J$ completes the definition of the channel: $\vec{J}=\vec{J_{c_i}}+\vec{\ell_i}+\vec{s_i}$, where $\vec{s_i}$ denotes the Rydberg electron spin. To any total energy $E<I_i$ corresponds an effective quantum number $\nu_i$ defined in an equation similar to Eq.~\eqref{eq:Energy_eq}:

\begin{equation}
E = I_i - \frac{R}{\nu_i^2}  \,
\label{eq:Energy_eqi}
\end{equation}

The channel wave function $\Psi_i(r,E)$ can again be described at $r>r_0$ as a linear combination of the Coulomb functions $f(\nu_i,\ell_i,r)$ and  $g(\nu_i,\ell_i,r)$. 

A particular eigenchannel $\alpha$ corresponds to a normal mode of the scattering from the doubly charged ionic core. Its wave function $\Psi_\alpha$ is an exact solution of the doubly charged ion-electron-electron compound in the range $r_c<r <r_0$. It describes a particular combination of Coulomb waves incoming in all the collision channels and gaining in each channel at $r_0$ after reflection the same phase shift $\pi \mu_\alpha$, with $\mu_\alpha$ the eigenquantum defect. The solution $\Psi(E)$ of the Schr\"{o}dinger equation at the energy $E$ over the range $r_c<r<r_0$ can be written as a linear combination of the eigenchannel wave functions $\Psi_{\alpha}$. Using the transformation $U_{i\alpha}$ at $r=r_0$, $\Psi(E)$  is expanded in terms of the collision channel functions $\Psi_i$, an expansion which is valid beyond the $r_0$. For an energy $E$ below the first ionization limit, $\Psi(E)$ describes a physical level if it vanishes at $r\rightarrow\infty$  in the $N$ ionization channels. Applying this boundary condition leads to the following equation:

\begin{equation}
det\left|U_{i\alpha} sin[\pi(\nu_i+\mu_{\alpha})]\right|=0.
\label{eq:Criterion}
\end{equation}

It is customary to ascribe $LS$ coupling scheme to the $\alpha$ channels, the electrostatic interaction being stronger at short distance than the spin-orbit interaction of the valence and Rydberg electrons, meanwhile the collision channels are described in $jj$ coupling scheme $n_0\ell_0j_0\,nlj\,J$. Intermediate eigenchannels $\overline{\alpha}$ defined in pure $LS$ coupling $n_0\ell_0\,nl\,SLJ$ are introduced together with the analytically known unitary matrix $U_{i{\overline{\alpha} }}$. The $U_{i\alpha}$ matrix is then factored into the product $U_{i{\overline{\alpha} }} V_{{\overline{\alpha}}{{\alpha}}}$. Due to its bielectronic character and its symmetry properties, the electrostatic interaction couples $\overline \alpha$ eigenchannels differing at most by the quantum numbers $n\ell$ of two electrons and associated with the same $L$ and $S$ values. $V_{{\overline{\alpha}}{{\alpha}}}$ can thus be written in terms of a product of a few $2\times2$ rotation matrices $R_{{\alpha}{\overline{\alpha}}}(\theta_{{{\alpha}}{\overline{\alpha}}})$ with the angle $\theta_{{{\alpha}}{\overline{\alpha}}}$ considered to be a free parameter.

The problem is now fully parametrized with a small number of MQDT parameters ($\mu_\alpha$ and $\theta_{{{\alpha}}{\overline{\alpha}}}$), the later together with the known $U_{i{\overline{\alpha} }}$ determining $U_{i{\alpha}}$. Combining Eq. (\ref{eq:Criterion}) with Eq. (\ref{eq:Energy_eqi}) finally determines the level energies associated to the given set of parameters. The principal aim of the MQDT analysis is to determine, using the experimental energies of bound levels with the same parity $\Pi$ and total angular momentum  $J$, the optimum set of parameters ($\mu_\alpha$ and $\theta_{{{\alpha}}{\overline{\alpha}}}$) predicting level energies as close as possible to all experimental points. We define an error function $\chi^2$ which sums all the energy differences between experimental results and model energies, normalized both with the assumed accuracy for each data and by the total number of data points such that a converged fit should lead to $\chi^2\leq1$ \cite{Vaillant_2014}.

To present graphically the results, we use a Lu-Fano representation \cite{Lu_1970} which presents the variations of $-\nu_{1}(E)$ modulo 1 for the Rydberg channel as a function of the effective quantum number $\nu_{j}(E)$ of one of the perturbing channels with a different ionization limit $I_j$. Using Eq. (\ref{eq:Energy_eqi}), one can define the function $\nu_{k}=f(\nu_j)$ of any other channel $k$ converging toward the ionization limit $I_k$ as a function of $\nu_j$ at a fixed energy $E$:

\begin{equation}
\nu_{k}=\left(\frac{I_k-I_j}{R}+\frac{1}{\nu_j^2}\right)^{-1/2}.
\label{eq:Energy_eqi2}
\end{equation}

Applying Eq. (\ref{eq:Energy_eqi2}) to the Rydberg channel $k=1$ leads in the Lu-Fano plane $(-\nu_1,\nu_j)$ to a curve $\mathcal L$ consisting of a set of nearly straight vertical lines. Equation (\ref{eq:Criterion}) defines another set of curves $\mathcal S$ in the plane $(-\nu_1,\nu_j)$ with a number equal to the number of Rydberg series converging towards the threshold $I_1$. The intersection of $\mathcal L$ and $\mathcal S$ define graphically the predicted energies. Traditionally, one plots only $\mathcal S$ together with the experimental $(-\nu_1,\nu_j)$ points in order to compare them with theory.

More specifically, in our case with Ytterbium atoms, we use for the Rydberg constant $R_\text{Yb}=R_{\infty}*\frac{m_\text{Yb}}{m_\text{Yb}+m_\text{e}}$, with $R_{\infty}=10973731.5685$ m$^{-1}$ the Rydberg constant, $m_\text{Yb}$ the mass of the considered isotope and $m_\text{e}$ the mass of an electron. For Ytterbium 174, $m_\text{Yb}=173.93886$ uma and we find $R=10973696.959$ m$^{-1}$. Finally, unless specified otherwise, we used the MQDT models introduced and described in detail in \cite{Aymar_1980}: We consider each perturbing level to be the lowest member of a new Rydberg series such that the channel eigenquantum defect $\mu_\alpha$ is redundant with the ionization limit which is thus kept constant. The only free-parameter ionization limit is the first ionization limit.

\section{Results : $\mathbf{6sns ~ {}^1S_0}$ series}
\label{sec:results}
\subsection{$6sns$ to $6s(n+1)s$ microwave transitions}

We first present the data for the microwave transitions $6sns^1S_0\rightarrow6s(n+1)s^1S_0$ that have been measured at UVA with principal quantum number ranging from $n=34$ to $n=39$ plus $n=42$. These data complement previous measurements \cite{Maeda_1992}, providing a larger set from $n=34$ to $n=52$. We stop at $n=52$ because the $n=53$ value in \cite{Maeda_1992} appears to be in error. We report in Table \ref{6sns_MW} the combined data, mentioning the observed statistical uncertainty of our measurements. For all the microwave transitions we present the observed frequencies, which are half the frequency intervals between the states. We compare the data set to the theoretical MQDT prediction computed thanks to these data and laser spectroscopy data presented later. Finally, Table \ref{6sns_MW} also presents the difference between the two. We then observe a standard deviation of around 0.15 MHz on the frequencies, leading thus to around 0.3 MHz uncertainty in the energy intervals, compatible with the expected accuracy. We note the two measurements at $n=39$ and $n=40$ which seem to deviate from the others. This could be the result of the transition between the two data sets with different average offsets, or to the effect of an unidentified weak perturber at around $n=40$. Nevertheless, this deviation is still compatible with our uncertainties and it is therefore impossible to decide if it is simply experimental error or a weak perturbation.

\begin{table}[htb]
\begin{ruledtabular}
\begin{tabular}{cccc}
\multicolumn{1}{c}{n} & \multicolumn{1}{c}{$f_{Exp}$ (MHz)} & \multicolumn{1}{c}{$f_{Th}$ (MHz)} & \multicolumn{1}{c}{$f_{Th}-f_{Exp}$ (MHz)} \\ \hline
\rule{0pt}{10pt}34 & 119119.66(6) & 119119.73 & 0.07\\
35 & 108044.35(7) & 108044.35 & 0.00\\
36 & 98299.99(6) & 98299.91 & -0.08\\
37 & 89692.36(6) & 89692.43 & 0.07\\
38 & 82060.74(6) & 82060.93 & 0.19\\
39 & 75270.32(5) & 75270.92 & 0.60\\
40 & 69210.18 & 69209.62 & -0.56\\
41 & 63781.70 & 63781.88 & 0.18\\
42 & 58907.08(5) & 58907.03 & -0.05\\
43 & 54516.42 & 54516.48 & 0.06\\
44 & 50551.50 & 50551.54 & 0.04\\
45 & 46962.06 & 46961.89 & -0.17\\
46 & 43704.31 & 43704.18 & -0.13\\
47 & 40740.96  & 40740.88 & -0.08\\
48 & 38039.57 & 38039.53 & -0.04\\
49 & 35571.81 & 35571.75 & -0.06\\
50 & 33312.97 & 33312.89 & -0.08\\
51 & 31241.37 & 31241.30 & -0.07\\
52 & 29337.83 & 29337.94 & 0.11\\
\end{tabular}
\end{ruledtabular}
\caption{$6sns$-$6s(n+1)s$ two-photon frequencies. New measurements for $n=34-39$ and $n=42$ are provided with the observed statistical uncertainty. Other data are from \cite{Maeda_1992}. $f_{Exp}$ are the experimentally measured frequencies, $f_{Th}$ are the theoretical frequencies extracted from the MQDT analysis presented later and the last column presents their difference $f_{Th}-f_{Exp}$.}
\label{6sns_MW}
\end{table}

\subsection{$6sns$ laser spectroscopy}

\begin{figure}[htb]
\mbox{
\centering
\includegraphics*[width=0.95\columnwidth]{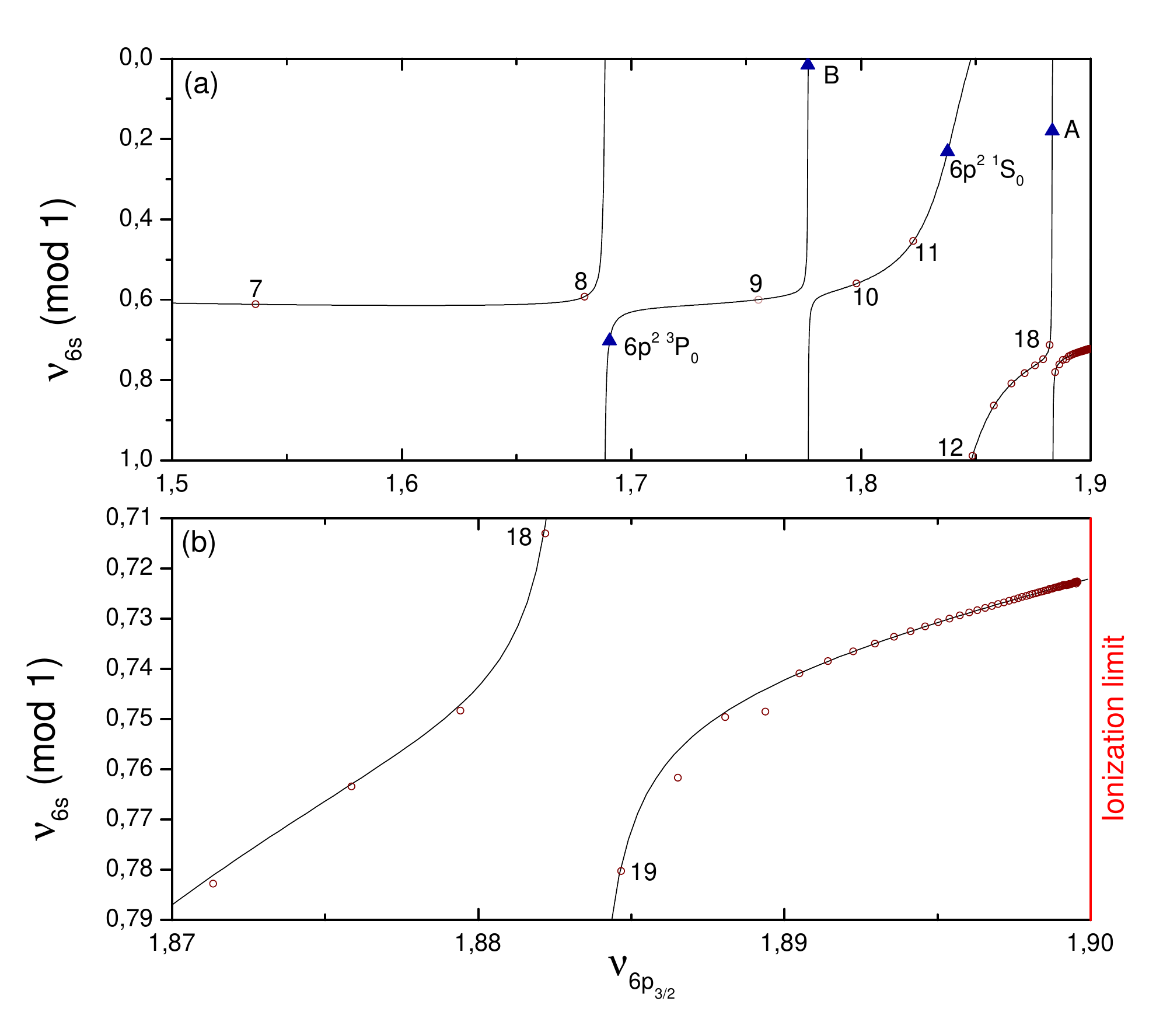}
	}
		\caption{Lu-Fano plot for the $6sns \,{}^{1}\!S_0$ series. The open circles represent the experimental positions of the $6sns$ bound states, the triangles those of the perturbers. $\nu_{6p_{3/2}}$ is calculated relative to $I_{6p_{3/2}}$ corresponding to the third channel. (a) - Evolution of $\nu_{6s}$ as a function of $\nu_{6p_{3/2}}$ for the whole energy range. We add the $n=9$ computed energy as a guide to the eye. (b) - $\nu_{6s}$ evolution above the $6s15s \,{}^{1}\!S_0$ level.}
\label{fig_LuFano_S}
\end{figure}

At LAC we have performed laser spectroscopic measurements of the energies of the $6sns\,{}^1\!S_0$ states in order to obtain their absolute energies. We can excite levels from $n=23$ to $n=80$, and the results can be found at the end in Table \ref{6sns_values}. In this energy range, no new perturbing state was identified. Nevertheless, in order to obtain an accurate fit over this broad energy range and to extract the ionization limit, we use an MQDT analysis. Indeed, as can be seen in Fig. (\ref{fig_LuFano_S}) presenting the Lu-Fano plot of this series, an important energy dependence influences the quantum defect due to low lying perturbing states \cite{Aymar_1980}. Therefore, the ionization limit fitted from a Ritz formula would be flawed.

To perform our MQDT analysis, we combined our laser and microwave spectroscopy data with the low lying levels observed in \cite{Aymar_1980}. A preliminary fit of data from \cite{Aymar_1980} demonstrated a standard deviation on levels from $n=7$ to $n=22$ of around 3 GHz and an average shift relative to our results for $n>22$ of around 3 GHz which we finally compensate by shifting the low lying energy levels by 0.1 cm$^{-1}$ (3 GHz). Wherever possible, it is beneficial to use the microwave results to compute the energy of one Rydberg level from the previous or the next. Doing so, the microwave errors accumulate and induce a residual random drift which is minimized if the starting level is chosen close to the center of the microwave data set. The drift is estimated to 0.6 MHz when starting from the $n=42$ level which was found the closest to predictions of another preliminary fit of the laser spectroscopy data only. The obtained combined data used in our final MQDT model is also presented in Table \ref{6sns_values}.

\subsection{$6sns$ MQDT analysis}

\begin{table*}[t]
\caption{First six-channel model MQDT parameters for the $6sns \,{}^{1}S_0$  series of ytterbium.}
\label{6sns_MQDT_S}
\begin{ruledtabular}
\begin{tabular}{lllllll}
i, $\bar{\alpha}$, $\alpha$ & 1 & 2 & 3 & 4 & 5 & 6 \\
\hline
\rule{0pt}{10pt}$\Ket{i}$ & $6s_{1/2}ns_{1/2}$ & $4f^{13}5d6s6p \,\, \mathbf{a}$ & $6p_{3/2}np_{3/2}$ & $4f^{13}5d6s6p \, \mathbf{b}$ & $6p_{1/2}np_{1/2}$ & $4f^{13}5d6s6p \, \mathbf{c}$ \\
$I_i$ & 50443.070417 & 83967.7 & 80835.39 & 83967.7 & 77504.98 & 83967.7 \\
$\Ket{\bar{\alpha}}$ & $6sns\,{}^1\!S_0$ & $4f^{13}5d6s6p \,\, \mathbf{a}$ & $6pnp\,{}^1\!S_0$ & $4f^{13}5d6s6p \, \mathbf{b}$ & $6pnp\,{}^3\!P_0$ & $4f^{13}5d6s6p \, \mathbf{c}$ \\
${\mu_{\alpha}}^{\!\!\!\!0}$ & 0.3551533 & 0.2045376 & 0.1163786 & 0.2954456 & 0.2576527 & 0.1560523 \\
${\mu_{\alpha}}^{\!\!\!\!1}$ & 0.2772279 & 0.0 & 0.0 & 0.0 & 0.0 & 0.0 \\
$V_{\bar{\alpha}\alpha}$ & $\theta_{12}$=0.1265847 & $\theta_{13}$=0.3001512 & $\theta_{14}$=0.05671288 & $\theta_{34}$=0.1142027 & $\theta_{35}$=0.09858364 & $\theta_{16}$=0.1419855 \\ \\
\multirow{6}{*}{$U_{i\alpha}$} & 0.9366039 & -0.1262469 & -0.296065 & -0.01994239 & 0.02928208 & -0.1338851 \\
 & 0.119197 & 0.9919989 & -0.03767875 & -0.002537971 & 0.003726589 & -0.01703891 \\
 & -0.2385956 & 0 & -0.7127722 & 0.1024783 & 0.6506634 & 0.03410663 \\
 & 0.05611209 & 0 & 0.113219 & 0.9918887 & -0.01119784 & -0.008021079 \\
 & 0.1687126 & 0 & 0.6245504 & -0.0724631 & 0.7587097 & -0.02411703 \\
 & 0.1415089 & 0 & 0 & 0 & 0 & 0.989937 \\
\end{tabular}
\end{ruledtabular}
\end{table*}

\begin{table*}[t]
\caption{Second six-channel model MQDT parameters for the $6sns \,{}^{1}S_0$  series of ytterbium.}
\label{6sns_MQDT_T}
\begin{ruledtabular}
\begin{tabular}{lllllll}
i, $\bar{\alpha}$, $\alpha$ & 1 & 2 & 3 & 4 & 5 & 6 \\
\hline
\rule{0pt}{10pt}$\Ket{i}$ & $6s_{1/2}ns_{1/2}$ & $4f^{13}5d6s6p \,\, \mathbf{a}$ & $6p_{3/2}np_{3/2}$ & $4f^{13}5d6s6p \, \mathbf{b}$ & $6p_{1/2}np_{1/2}$ & $4f^{13}5d6s6p \, \mathbf{c}$ \\
$I_i$ & 50443.070425 & 83967.7 & 80835.39 & 83967.7 & 77504.98 & 83967.7 \\
$\Ket{\bar{\alpha}}$ & $6sns\,{}^1\!S_0$ & $4f^{13}5d6s6p \,\, \mathbf{a}$ & $6pnp\,{}^1\!S_0$ & $4f^{13}5d6s6p \, \mathbf{b}$ & $6pnp\,{}^3\!P_0$ & $4f^{13}5d6s6p \, \mathbf{c}$ \\
${\mu_{\alpha}}^{\!\!\!\!0}$ & 0.3546985 & 0.2045358 & 0.1091799 & 0.2967248 & 0.3140328 & 0.08932482 \\
${\mu_{\alpha}}^{\!\!\!\!1}$ & 0.2715197 & 0.0 & 0.0 & 0.0 & 0.0 & 0.0 \\
$V_{\bar{\alpha}\alpha}$ & $\theta_{12}$=0.126671 & $\theta_{13}$=0.3189996 & $\theta_{14}$=0.01177203 & $\theta_{34}$=0.07887204 & $\theta_{35}$=-0.009607895 & $\theta_{56}$=0.9083208 \\ \\
\multirow{6}{*}{$U_{i\alpha}$} & 0.9418765 & -0.1263325 & -0.3109963 & 0.01345814 & -0.0018379 & 0.002356042 \\
 & 0.1199507 & 0.9919879 & -0.03960628 & 0.001713933 & -0.0002340618 & 0.0003000488 \\
 & -0.2560493 & 0 & -0.7781676 & 0.06409142 & 0.3505288 & -0.4493503 \\
 & 0.01177176 & 0 & 0.07878119 & 0.9968221 & 0.0004655745 & -0.0005968297 \\
 & 0.1810542 & 0 & 0.5384806 & -0.04531948 & 0.5054085 & -0.6478938 \\
 & 0 & 0 & 0 & 0 & 0.788472 & 0.6150707 \\
\end{tabular}
\end{ruledtabular}
\end{table*}

We first tried the MQDT model from \cite{Aymar_1980} which is based on a 5 channel model associated with the following four perturbing levels: the $6p^{\,2} \, {}^3P_0$ (42436.94 cm$^{-1}$), the $4f^{13}5d6s6p ~ B$ (46081.54 cm$^{-1}$), the $6p^{\,2} \, {}^1S_0$ (48344.38 cm$^{-1}$) and the $4f^{13}5d6s6p ~ A$ (49897.32 cm$^{-1}$). The analysis uses the known $LS-jj$ transformation matrix within the $6p^{\,2} \, J=0$ levels:

\[
U_{i\overline{\alpha}}=\left\{\begin{array}{ccccc}
  1   &   0  & 0 & 0 &0 \\
	 0   &   1  & 0 & 0 &0 \\
	 0   &   0  &-\sqrt{2/3} & 0 &\sqrt{1/3} \\
	0   &   0  & 0 & 1 &0 \\
	0   &   0  & \sqrt{1/3} & 0 &\sqrt{2/3}
			\end{array}
			\right\}.
			\]

We have found inconsistencies with previous results using this model. The newly found optimum parameters could not reproduce the previous fit quality for $n<23$, showed a clear residual differential drift for $n>22$ levels and had a minimum $\chi^2$ of around 60 while considering an uncertainty of 3 MHz on all new data. We have thus looked for possible improvement of the model. Within the $4f^{13}5d6s6p$ configuration, the four $4f^{13} (^{2}F_{7/2}) 5d6s6p, J=0$ levels are the lowest in energy and might perturb the Rydberg series. The two lowest levels should have a dominant ${}^{3}\!P_0$ character and are already included as $A$ and $B$ in the original model. The two others should consist of a mixture of ${}^{1}\!S_0$ and ${}^{3}\!P_0$ with a slightly prevailing ${}^{1}\!S_0$ character. A narrow auto-ionized level has been observed by two photon spectroscopy from the ground state at the energy of 51842.42 cm$^{-1}$ and has been identified as a $J=0$ level  \cite{Camus_1980}. Moreover, an analysis of the configuration mixing in the even $J=0$  spectrum has pointed out the strong configuration mixing between $4f^{13} 5d6s6p$ and $4f^{14} 6p^2$ and has predicted a level with substantial $^{1}\!S_0$ character at a similar energy of 51749 cm$^{-1}$ \cite{Wyart_1979}. We thus suspect the influence of a perturber above the first ionization limit. But due to the absence of data on the fourth expected level and the fact that we do not know which one might explain the observed drift, we consider two possible options for a new 6 channel model with either a new ${}^{1}\!S_0$ channel directly coupled to the $6sns \,{}^{1}\!S_0$ Rydberg series or an indirect coupling with a new ${}^{3}\!P_0$ channel coupled to the $6p^{\,2} \, {}^3P_0$ channel. We then adapt $V_{{\overline{\alpha}}{\alpha}}$ by multiplying it on the right with the rotation matrix $R_{16}(\theta_{16})$ or $R_{56}(\theta_{56})$ coupling the sixth channel labeled {\bf{c}} (corresponding to the additional perturbing level $C$). We also have to extend $U_{i\overline{\alpha}}$. Following \cite{Aymar_1980} we consider $\overline{\alpha}=i$ for this $4f^{13}5d6s6p$ sixth channel.

We manage to find a good agreement over the whole data set with both models, obtaining a $\chi^2$ value of around 0.81 with uncertainties set to 3 GHz for $n<23$, 3 MHz for laser spectroscopy and 0.6 MHz for microwave spectroscopy data. The resulting model optimum parameters are presented in Tables \ref{6sns_MQDT_S} and \ref{6sns_MQDT_T}. In the indirect coupling model, the rotation angle $\theta_{56}$ is as expected much higher than the rotation angle $\theta_{16}$ of the direct coupling model. We compare the new optimum sets of our 6-channel model parameters to the set of the previous 5-channel model (Table 1 of Ref.~\cite{Aymar_1980}): The addition of channel {\bf{c}} reduces the mixing between the Rydberg channel and the $4f^{13}5d6s6p$ {\bf b} channel with a prevailing $^3\!P_0$ character \cite{Wyart_1979}. Simultaneously there is a redistribution in the mixing between the $6pnp\, {}^1\!S_0$ channel and both $4f^{13}5d6s6p$ {\bf b} and $6pnp\, {}^3\!P_0$ channels, the first one increasing significantly and the second one decreasing. A change in channel mixing also appears in a larger energy variation $\mu^1_\alpha$ in the eigenchannel quantum defect of the Rydberg series. Since level $C$ is above the first ionization limit, it is auto-ionizing and its energy cannot be determined with Eq. \ref{eq:Criterion} which only applies to bound levels lying below the first limit. It has been shown \cite{Lee_1973, Seaton_1983, Lecomte_1987} that one can deduce the energy and the width of an auto-ionization resonance restricting calculations on closed channels. The first model finds an energy of 51591 cm$^{-1}$ and a total width of 97 cm$^{-1}$ similar to the second model which finds an energy of 51537 cm$^{-1}$ and a total width of 89 cm$^{-1}$. We thus find in both models a width significantly larger than the width $\Gamma=5$ cm$^{-1}$ of the level observed in \cite{Camus_1980}. Two possible explanations are as follows. First, an additional perturber with large coupling is indeed present, but its width precludes its observation as a distinct feature. Second, there are several additional perturbers. Both models demonstrate the necessity to introduce a $6^{th}$ channel in order to obtain a converged fit but the MQDT approach cannot directly provide identification of the perturbers above the first ionization limit and additional experimental data and reliable level identification are necessary to develop a more elaborate MQDT model. 

The two models predict similar energies, with differences well below our uncertainties and we choose to display the predicted level energies of first model in Table \ref{6sns_values}. In the next column, we can see that the residual difference between experimental measurements and the theoretical predictions are compatible with the expected uncertainty for each point of the data set, except for $n=40$ as discussed in the microwave section. The results are also displayed in a Lu-Fano representation in Fig. \ref{fig_LuFano_S}(a). Note that we added the computed $n=9$ level as a guide to the eye. The observation of this level could improve the model in this energy range. Fig. \ref{fig_LuFano_S}(b) focuses closer to the energy range of the new spectroscopic data to emphasize the accuracy improvement for $n>22$. As previously mentioned, a slow variation of the quantum defect is observed, mainly due to the wide energy range influence of the strong perturber $6p^{\,2} \, {}^1\!S_0$ but also from at least one perturbing level above the first ionization limit.

From this fit, we extract a new value for the first ionization limit of 50443.07042 cm$^{-1}$ or 1512.2452070 THz. Once the ionization limit is extracted, we can also extract the quantum defects for each level. As the quantum defect refers to a difference in energy, it is not sensitive to the absolute uncertainty of our measurement of 7.5 MHz but only to the observed reproducibility of 3 MHz. Moreover, the quantum defects displayed in Table \ref{6sns_values} are the ones predicted by the MQDT model which averages the measurement statistical noise. Evaluating the residual uncertainty on these quantum defects is too complex, but it should be below the initial noise of 3 MHz and probably below 1 MHz in the microwave data set range.

Finally, in order to provide the reader with a simple predictive tool, we perform a fit of the found quantum defects to a Ritz formula. To obtain a reliable fit on the whole new data set of $n>22$, we find that we need to use six parameters. We also test a simpler fit with only three parameters and find that we have to restrict to $n\geq34$ to maintain accuracy for all level energies and microwave transitions. The resulting Ritz parameters and validity range are given in Table \ref{6snl_Ritz}, together with parameters for the other series. Note that the fitted values of $\delta_0$ do not agree for the two different ranges. This justifies the use of MQDT to fit the ionization limit as the Ritz formula would thus probably lead to errors.

\section{Results : $\mathbf{6snd ~ {}^{3,1}D_2}$ series}
\subsection{$6s(n+1)s$ to $6snd$ microwave transitions}

Once the energy levels of one Rydberg series have been accurately measured, they can serve as a reference to perform microwave measurements towards higher $\ell$ levels. Although the $6snd \,{}^{1}D_2$ levels are optically accessible in both experiments, we will get a better accuracy for level energies with microwave transitions from $6s(n+1)s \,{}^{1}S_0$. We thus first present in Table \ref{6snsnd_MW} the corresponding microwave transition frequencies measured at UVA.

\begin{table}[htb]
\begin{ruledtabular}
\begin{tabular}{cccc}
\multicolumn{1}{c}{n} & \multicolumn{1}{c}{$f_{Exp}$ (MHz)} & \multicolumn{1}{c}{$f_{Th}$ (MHz)} & \multicolumn{1}{c}{$f_{Th}-f_{Exp}$ (MHz)} \\ \hline
\rule{0pt}{10pt}28  &117774.09(11) & 117774.17 & 0.08\\
29  &104637.58(11) & 104637.48 & -0.10 \\
30  & 93459.24(6) & 93459.29 & 0.05 \\
31  & 83840.26(8) & 83840.12 & -0.14 \\
32  & 75504.60(7) & 75504.63 & 0.03 \\
33  & 68241.11(5) & 68241.25 & 0.14 \\
34  & 61881.20(8) & 61881.27 & 0.07 \\
35  & 56288.03(9) & 56287.82 & -0.21 \\
36  & 51348.53(7) & 51348.47 & -0.06 \\
37  & 46970.16(5) & 46970.24 & 0.08 \\
38  & 43075.55(8) & 43075.57 & 0.02 \\
39  & 39599.77(7) & 39599.57 & -0.20 \\
40  & 36487.09 & 36487.43 & 0.33 \\
41  & 33692.83 & 33692.76 & -0.07 \\
42  & 31175.94(6) & 31176.15 & 0.21 \\
43  & 28903.64(5) & 28903.86 & 0.22 \\
44  & 26846.92 & 26846.97 & 0.05\\
45  & 24980.84 & 24980.57 & -0.27 \\
46  & 23283.35 & 23283.10 & -0.25 \\
47  & 21736.13 & 21735.91 & -0.22 \\
48  & 20322.53 & 20322.72 & 0.19 \\
\end{tabular}
\end{ruledtabular}
\caption{ $6s(n+1)s$-$6snd$ two-photon transition frequencies. New measurements for $n=28-39$ and $n=42-43$ are provided with the observed statistical uncertainty. Other data are from \cite{Maeda_1992}. $f_{Exp}$ are the experimentally measured frequencies, $f_{Th}$ are the theoretical frequencies extracted from the MQDT presented later, and the last column presents their difference $f_{Th}-f_{Exp}$.}
\label{6snsnd_MW}
\end{table}

The effective quantum defect difference between the two measured levels being smaller, transitions down to $n=28$ are now accessible with our microwave frequency range. We thus present data from $n=28$ to $n=39$, plus $n=42$ and $n=43$ to complete previous measurements \cite{Maeda_1992} to a set from $n=28$ to $n=48$. To evaluate the measurement stability, we again add in Table  \ref{6snsnd_MW} the frequencies obtained later from the MQDT fit and the resulting difference. This difference displays again a standard deviation in the measurement of around 0.15 MHz compatible with the expected accuracy with no exception.

\subsection{$6snd$ laser spectroscopy}

At LAC, we have completed the laser spectroscopy measurements of the $6snd \,J=2$ Rydberg series. Note that we detect the corresponding triplet state which is necessary for an MQDT analysis. The level energies from $n=23$ to $n=80$ have been measured, and the results are displayed in Table \ref{6snd_values}. We then combine again the laser spectroscopy data, the microwave spectroscopy data and $n<23$ data from \cite{Aymar_1980}. We find a quite large standard deviation of around 65 GHz for the $n<23$ data. Moreover, the microwave data are now connecting to the known $6sns$ series. The energy of $6snd ~ {}^{1}D_2$ levels from $n=28$ to $n=48$ can thus be computed from the $6s(n+1)s \,{}^{1}S_0$ energies predicted by the MQDT model, adding twice the transition frequency. We therefore assume an uncertainty of 0.3 MHz on these energies, although this neglects the average uncertainty of the MQDT fit of $6sns\,^{1}\!S_0$ itself.

\begin{figure}[htb]
\mbox{
\centering
\includegraphics*[width=0.95\columnwidth]{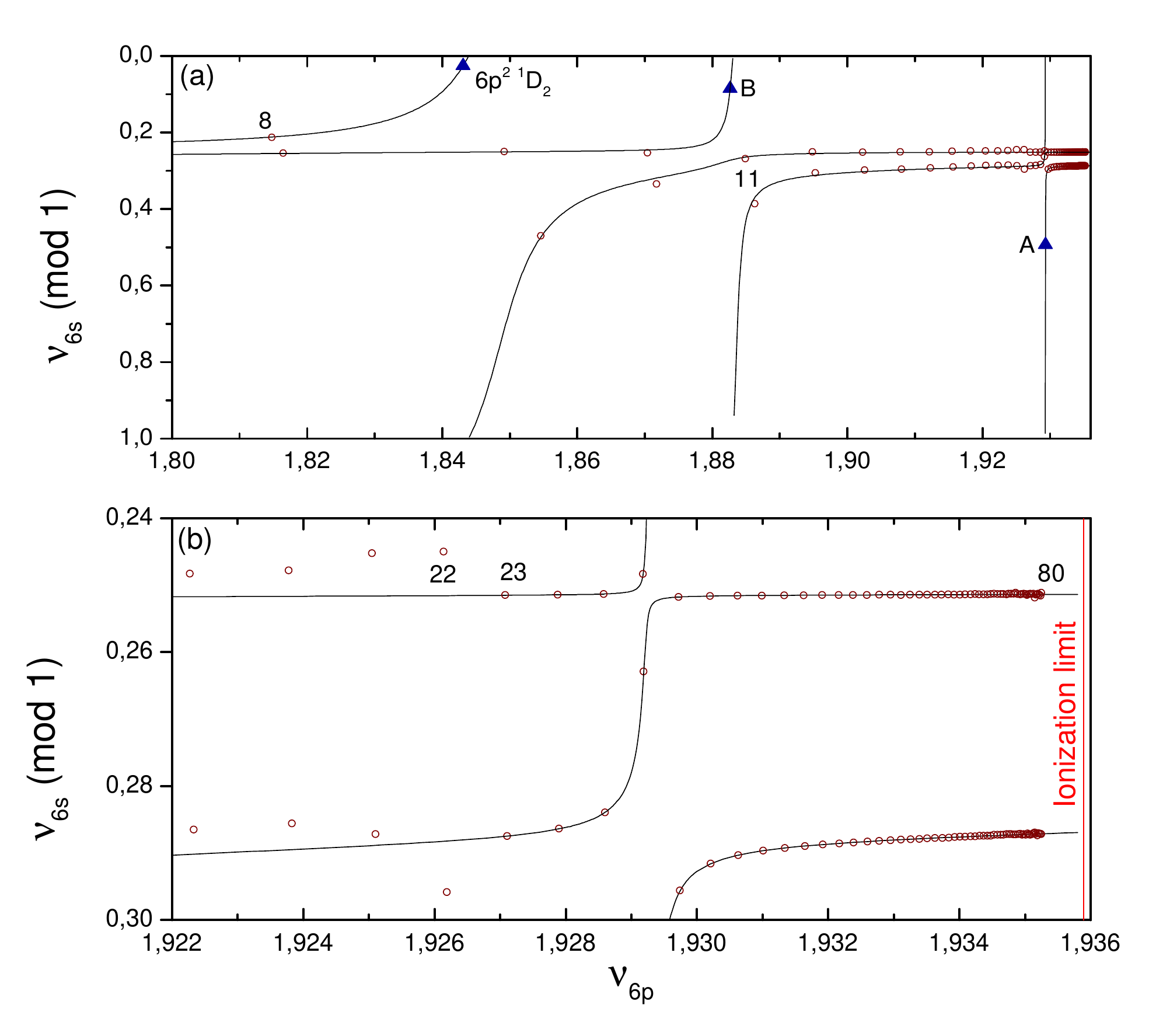}
	}
		\caption{Lu-Fano plot for the $6snd \,{}^{3,1}D_2$ series. The open circles represent the experimental positions of the $6snd$ bound states, the triangles those of the perturbers. $\nu_{6p}$ is calculated relative to $I_{6p}$ corresponding to the fifth channel. Plot (a) - $\nu_{6s}$ evolution for the whole energy range. Plot (b) - $\nu_{6s}$ evolution above the $6s19d$ levels.}
\label{fig_LuFano_D}
\end{figure}

\subsection{$6snd$ MQDT analysis}

\begin{table*}[t]
\caption{The five-channel MQDT parameters for the $6snd \, {}^{3,1}D_2$ series of ytterbium.}
\label{6snd_MQDT}
	\begin{ruledtabular}
		\begin{tabular}{llllll}
			i, $\bar{\alpha}$, $\alpha$ & 1 & 2 & 3 & 4 & 5 \\
			\hline
\rule{0pt}{10pt}$\Ket{i}$ & $6s_{1/2}nd_{5/2}$ & $6s_{1/2}nd_{3/2}$ & $4f^{13}5d6snp \,\, \mathbf{a}$ & $4f^{13}5d6snp \, \mathbf{b}$ & $6pnp$ \\
$I_i$ & 50443.0704 & 50443.0704 & 83967.7 & 83967.7 & 79725.35 \\
$\Ket{\bar{\alpha}}$ & $6snd\,{}^1D_2$ & $6snd\,{}^3D_2$ & $4f^{13}5d6snp \,\, \mathbf{a}$ & $4f^{13}5d6snp \, \mathbf{b}$ & $6p^2\,{}^1D_2$ \\
${\mu_{\alpha}}^{\!\!\!\!0}$ & 0.7295231 & 0.7522912 & 0.1961204 & 0.2336927 & 0.1528761 \\
${\mu_{\alpha}}^{\!\!\!\!1}$ & -0.02289746 & 0.09094972 & 0.0 & 0.0 & 0.0 \\
$V_{\bar{\alpha}\alpha}$ & $\theta_{13}$=0.005224693 & $\theta_{14}$=0.03972731 & $\theta_{24}$=-0.007083118 & $\theta_{15}$=0.1049613 & $\theta_{25}$=0.07219257 \\ \\
\multirow{5}{*}{$U_{i\alpha}$} & 0.7697155 & 0.6251608 & -0.004047011 & -0.02628364 & -0.1265106 \\
 & -0.62847 & 0.7771578 & 0.003304371 & 0.03060469 & 0.01017855 \\
 & 0.005191816 & -3.798545e-005 & 0.9999864 & -0.0002075022 & -0.000545631 \\
 & 0.03949828 & -0.007359173 & 0 & 0.9991859 & -0.003639746 \\
 & 0.1047687 & 0.07173292 & 0 & 0 & 0.9919062 \\
\end{tabular}
\end{ruledtabular}
\end{table*}

For this Rydberg series, the 5 channel MQDT model from \cite{Aymar_1980} leads to good agreement, with the expected uncertainties. It involves the three perturbing levels: $6p^{\,2} \, {}^1D_2$ (47420.97 cm$^{-1}$), $4f^{13}5d6s6p ~ B$ (48762.52 cm$^{-1}$) and $4f^{13}5d6s6p ~ A$ (50244.38 cm$^{-1}$) and the known $LS-jj$ transformation matrix within the $6snd \, J=2$ levels:

\[
U_{i\overline{\alpha}}=\left\{\begin{array}{ccccc}
 \sqrt{3/5}   &  \sqrt{2/5}  & 0 & 0 &0 \\
	-\sqrt{2/5}   & \sqrt{3/5} & 0 & 0 &0 \\
	 0   &   0  &1 & 0 &0 \\
	0   &   0  & 0 & 1 &0 \\
	0   &   0  & 0 & 0 &1
			\end{array}
			\right\}
			\]

The optimum set of our 5-channel model (Table \ref{6snd_MQDT}) can be compared to the set of the same 5-channel model reported in Table 4 of Ref.~\cite{Aymar_1980}. The two sets are very similar, except the energy variation of the eigenquantum defect in the $6snd \,{}^{3}D_2$ series is strongly reduced in our description. This is due to the introduction in our fit of the energy of the triplet levels with $62 \le n \le 80$ and to the increase in the accuracy of our measurements.

The fitted energies are also presented in Table \ref{6snd_values}. We once again extract a value for the first ionization limit: 50443.07040 cm$^{-1}$ or 1512.2452064 THz which is compatible with the one found for the $6sns \,{}^{1}S_0$ series with only 600 kHz difference, well below our uncertainty. This confirms our measurement stability and the completeness of the models of both series. The obtained MQDT parameters are presented in Table \ref{6snd_MQDT} and results are presented in Fig. \ref{fig_LuFano_D} using a Lu-Fano representation.

Like for the $6sns \,{}^{1}\!S_0$ series, we perform a final fit with a Ritz formula on the $6snd \,{}^{3,1}\!D_2$ series. Due to the perturber around $n=26$, it is not possible to apply such a fit for the whole new data set, although the perturber is weakly coupled. On the singlet series which is the most coupled, we find a valid fit with 6 parameters for $n\geq 31$ and to reduce to only 3 parameters, one has to restrict the fit to $n\geq40$. The triplet series is less coupled, and we could find a correct fit with 6 parameters starting at $n=28$. To reduce to only 3 parameters, the range needs to be restricted to $n\geq35$. The resulting optimum parameters are displayed in Table \ref{6snl_Ritz}.

\begin{table*}[t]
\caption{Fitted Ritz parameters and validity range compatible with our experimental uncertainty. The fit is performed using the found average first ionization limit of 50443.07041 cm$^{-1}$. For each series we first fit the largest possible data range with coefficients up to $\delta_5$ and perform next a fit on a restricted data range in order to maintain the accuracy with only 3 parameters.}
\label{6snl_Ritz}
\begin{ruledtabular}
\begin{tabular}{llllllll}
Series & Validity range & $\delta_0$ &  $\delta_1$ & $\delta_2$ & $\delta_3$ & $\delta_4$ & $\delta_5$ \\
\hline

$6sns \, {}^1 \! S_0$ & $23\leq n\leq 80$ & 4.278337 & -5.625 & 91.65 & -156050 & -49725000 & 11021000000 \\
$6sns \, {}^1 \! S_0$ & $34\leq n\leq 80$ & 4.278312 & -5.4898 & -320.02 & $\,$ & $\,$ & $\,$ \\
$6snd \, {}^1 \! D_2$ & $31\leq n\leq 80$ & 2.713094 & -1.8646 & -2145.5 & 3940500 & -3103600000 & 1069000000000 \\
$6snd \, {}^1 \! D_2$ & $40\leq n\leq 80$ & 2.713011 & -1.1047 & -373.87 & $\,$ & $\,$ & $\,$ \\
$6snd \, {}^3 \! D_2$ & $28\leq n\leq 80$ & 2.748679 & -0.5200 & -1186.01 & 1564600 & -981340000 & 242600000000 \\
$6snd \, {}^3 \! D_2$ & $35\leq n\leq 80$ & 2.748627 & -0.1034 & -2.10 & $\,$ & $\,$ & $\,$ \\
\end{tabular}
\end{ruledtabular}
\end{table*}

\subsection{$6snd$ to $6s(n+1)d$ microwave transitions}

As a consistency check, we have measured the $6snd-6s(n+1)d$ transitions. Within the available microwave frequency range we could only observe transitions for $n\geq32$. We have performed the measurements for $32\leq n\leq 38$ which complement the previous observations in \cite{Maeda_1992} and the resulting data set is given in Table \ref{6snd_MW}. We also present the prediction from the MQDT model and the difference between the two.

\begin{table}[htb]
\begin{ruledtabular}
\begin{tabular}{cccc}
\multicolumn{1}{c}{n} & \multicolumn{1}{c}{$f_{Exp}$ (MHz)} & \multicolumn{1}{c}{$f_{Th}$ (MHz)} & \multicolumn{1}{c}{$f_{Th}-f_{Exp}$ (MHz)} \\ \hline
\rule{0pt}{10pt}32 & 124498.32 (14)	& 124497.60 & -0.72\\
33 & 112759.83 (6)	& 112759.75 &	-0.08\\
34 & 102451.05 (6)	& 102450.89 &	-0.16\\
35 & 93360.74 (6) & 93360.56	& -0.18\\
36 & 85314.28 (7) &	85314.20	& -0.08\\
37 & 78166.60 (7) &	78166.27 &	-0.33\\
38 & 71794.73 (4) &	71794.91 & 0.18\\
39 & 66097.45  &	66097.48 & 0.03\\
40 & -  &	60987.21 & -\\
41 & 56390.46  &	56390.42 & -0.04\\
42 & 52244.45  &	52244.19 & -0.26\\
43 & 48494.83  &	48494.65 & -0.18\\
44 & 45095.63  &	45095.49 & -0.14\\
45 & 42006.90  &	42006.71 & -0.19\\
46 & 39193.92  &	39193.70 & -0.22\\
47 & 36626.62  &	36626.33 & -0.29\\
48 & 34278.59  &	34278.40 & -0.19\\
49 & 32127.16  &	32126.93 & -0.23\\
50 & - & 30151.81 & -\\
51 & 28335.49  &	28335.32 & -0.17\\
52 & 26662.12  &	26661.86 & -0.26\\
53 & 25117.98  &	25117.65 & -0.33\\
54 & 23690.72  &	23690.39 & -0.33\\
55 & 22369.66  &	22369.27 & -0.39\\
\end{tabular}
\end{ruledtabular}
\caption{$6snd$-$6s(n+1)d$ 2-photon transition frequencies. New measurements for $n=32-38$ are provided with the observed statistical uncertainty. Other data are from \cite{Maeda_1992}. $f_{Exp}$ are the experimentally measured frequencies, $f_{Th}$ are the theoretical frequencies extracted from the MQDT analysis presented later, and the last column presents their difference $f_{Th}-f_{Exp}$.}
\label{6snd_MW}
\end{table}

We find only small discrepancies with typically around 200 kHz difference. We also notice the non vanishing average difference, with the theoretical frequencies being generally smaller than the measured ones. This could be attributed to residual drift errors of the MQDT as well as the signature of a systematic shift in the microwave measurements, but all compatible with our uncertainties.

\section{Discussion}

The precise knowledge of the ytterbium spectra is motivated by its use as a Rydberg interacting gas and as the ion in a clock. Using high resolution laser spectroscopy, we have measured the term energies of the $6sns \,{}^1\!S_0$ and $6snd \,{}^{3,1}\!D_2$ Rydberg series over $23\leq n \leq 80$. The absolute accuracy is expected to be 7.5 MHz at one standard deviation and the observed statistical uncertainties are 2.5 MHz. We note that they are significantly lower than the expected total of 5 MHz. Each measurement corresponds to a fit of the laser excitation line over several experiments and the wavemeter's statistical measurement noise is averaged down. Complementary microwave transitions have been measured for the $6sns\,^1\!S_0$ - $6s(n+1)s\,^1\!S_0$ intervals from $n=34$, the $6snd\,^1\!D_2$ - $6s(n+1)d\,^1\!D_2$ intervals from $n=32$, and the $6s(n+1)s\,^1\!S_0$- $6snd\,^1\!D_2$ intervals from $n=28$, each set complementing previous measurements. We observe typical statistical uncertainties of 0.3 MHz on level differential energies and can thus refine the optical measurements.

The combined results are fit to a MQDT model, which characterizes these perturbed series with only a few parameters. These analyses provide an average ionization limit  I$_{6s}=50443.07041(25)$ cm$^{-1}$, compatible with the previously accepted value and an improved accuracy of more than 2 orders of magnitude. The inferred quantum defects correspond to a relative energy uncertainty smaller than 2.5 MHz in general, and probably smaller than 1 MHz in the microwave data range.

Although MQDT analysis of molecular spectra \cite{Osterwalder_2004} or rare gas Rydberg series \cite{Lee_1973} sometimes uses perturbing channels with only auto-ionizing levels, it is not common for other Rydberg atoms and it is the first time for Ytterbium. In the future, it will be possible to complete the laser spectroscopy with a three-photon excitation through $6s \, 6p ~ {}^3 \! P_1 $ and $6s \, 6d ~ {}^{3,1}\!D_2$ to $6snp$ and $6snf$ singlet or triplet levels. The precise results obtained for $6snd$ will also allow performing two-photon microwave spectroscopy towards $6sng$ levels and from these to further high $\ell$ levels in the prospect of a precise calculation of the ionic core polarizability.

\section{Acknowledgements}

We are grateful to F. Merkt for discussions on MQDT. T.\~F. G. acknowledges support from ENS Cachan and P. C. acknowledges support from University of Virginia for mutual visits. This work was supported by ANR program COCORYM (ANR-12-BS04-0013), the public grant CYRAQS from Labex PALM (ANR-10-LABX-0039) (P.C.), the EU H2020 FET Proactive project RySQ (grant N. 640378), and the U. S. Department of Energy, Office of Science, Office of Basic Energy Sciences under Award Number DE-FG02-97ER14786.

\bigbreak

\begin{longtable*}{@{\extracolsep{\fill}}cccccc}
\caption{Experimental and theoretical energies of the $6sns \,{}^{1}S_0$ Rydberg series of $^{174}$Yb. $E_{LSp}$ is the new laser spectroscopy data. $E_{Cb}$ is the chosen combination of experimental data to be analyzed with MQDT: Data followed with $^{\blacktriangle}$ is taken from \cite{Aymar_1980} minus the observed shift of 0.1 cm$^{-1}$, data followed with $^{\bullet}$ is extrapolated from $E_{LSp}$ at $n=40$ with the microwave data presented in Table \ref{6sns_MW}, data without sign is from $E_{LSp}$. $E_{Th}$ is the theoretical energy obtained with our MQDT model presented in Table \ref{6sns_MQDT_S}. The next column corresponds to the difference between theoretical and experimental energies expressed in MHz and the last column is the quantum defect inferred from the MQDT model, computed assuming the ionization limit at 50443.07042 cm$^{-1}$.} \label{6sns_values} \\

\hline \hline  \\[-2.0ex] Assignment & $E_{new}$ (cm$^{-1}$) & $E_{comb.}$ (cm$^{-1}$) & $E_{th}$ (cm$^{-1}$) & $E_{exp}-E_{th}$ (MHz) & Q. Defect \\ \hline \\[-2.0ex]
\endfirsthead

\multicolumn{6}{c}%
{{\bfseries \tablename\ \thetable{} -- continued from previous page}} \\
\hline  \\[-2.0ex] \multicolumn{1}{c}{Assignment} &
\multicolumn{1}{c}{$E_{new}$ (cm$^{-1}$)} &
\multicolumn{1}{c}{$E_{comb.}$ (cm$^{-1}$)} &
\multicolumn{1}{c}{$E_{th}$ (cm$^{-1}$)} &
\multicolumn{1}{c}{$E_{exp}-E_{th}$ (MHz)} &
\multicolumn{1}{c}{Q. Defect} \\ \hline  \\[-2.0ex]
\endhead

\hline \multicolumn{6}{|r|}{{\rule{0pt}{10pt} Continued on next page}} \\ \hline
\endfoot

\hline \hline
\endlastfoot

$6s7s \, {}^1 \! S_0$ & $\,$ & 34350.55$^{\blacktriangle}$ & 34350.549906 & 2 & 4.388654 \\ 
$6s8s \, {}^1 \! S_0$ & $\,$ & 41939.78$^{\blacktriangle}$ & 41939.780040 & -2 & 4.407614 \\ 
$6p^2 \, {}^3 \! P_0$ & $\,$ & 42436.84$^{\blacktriangle}$ & 42436.840006 & -1 & - \\ 
$4f^{13}5d6s6p \, B$ & $\,$ & 46081.44$^{\blacktriangle}$ & 46081.439988 & 0 & - \\ 
$6s10s \, {}^1 \! S_0$ & $\,$ & 46893.14$^{\blacktriangle}$ & 46893.135007 & 149 & 4.440109 \\ 
$6s11s \, {}^1 \! S_0$ & $\,$ & 47808.39$^{\blacktriangle}$ & 47808.426066 & -1082 & 4.546200 \\ 
$6p^2 \, {}^1 \! S_0$ & $\,$ & 48344.28$^{\blacktriangle}$ & 48344.222696 & 1717 & - \\ 
$6s12s \, {}^1 \! S_0$ & $\,$ & 48723.39$^{\blacktriangle}$ & 48723.442248 & -1567 & 4.011611 \\ 
$6s13s \, {}^1 \! S_0$ & $\,$ & 49046.33$^{\blacktriangle}$ & 49046.346545 & -497 & 4.136174 \\ 
$6s14s \, {}^1 \! S_0$ & $\,$ & 49302.44$^{\blacktriangle}$ & 49302.604742 & -4939 & 4.190760 \\ 
$6s15s \, {}^1 \! S_0$ & $\,$ & 49499.14$^{\blacktriangle}$ & 49498.902231 & 7128 & 4.219180 \\ 
$6s16s \, {}^1 \! S_0$ & $\,$ & 49649.95$^{\blacktriangle}$ & 49649.941036 & 268 & 4.237369 \\ 
$6s17s \, {}^1 \! S_0$ & $\,$ & 49767.75$^{\blacktriangle}$ & 49767.685616 & 1930 & 4.253210 \\ 
$6s18s \, {}^1 \! S_0$ & $\,$ & 49859.41$^{\blacktriangle}$ & 49859.328529 & 2442 & 4.289097 \\ 
$4f^{13}5d6s6p \, A$ & $\,$ & 49897.22$^{\blacktriangle}$ & 49897.230880 & -327 & - \\ 
$6s19s \, {}^1 \! S_0$ & $\,$ & 49940.64$^{\blacktriangle}$ & 49940.692904 & -1587 & 4.220444 \\ 
$6s20s \, {}^1 \! S_0$ & $\,$ & 50001.25$^{\blacktriangle}$ & 50001.051310 & 5956 & 4.243638 \\ 
$6s21s \, {}^1 \! S_0$ & $\,$ & 50051.82$^{\blacktriangle}$ & 50051.873470 & -1604 & 4.251391 \\ 
$6s22s \, {}^1 \! S_0$ & $\,$ & 50094.61$^{\blacktriangle}$ & 50094.534923 & 2250 & 4.255941 \\ 
$6s23s \, {}^1 \! S_0$ & 50130.625484 & 50130.625484 & 50130.625506 & -0.66 & 4.259130 \\ 
$6s24s \, {}^1 \! S_0$ & 50161.409314 & 50161.409314 & 50161.409264 & 1.50 & 4.261552 \\ 
$6s25s \, {}^1 \! S_0$ & 50187.870220 & 50187.870220 & 50187.870190 & 0.88 & 4.263476 \\ 
$6s26s \, {}^1 \! S_0$ & 50210.777667 & 50210.777667 & 50210.777658 & 0.27 & 4.265050 \\ 
$6s27s \, {}^1 \! S_0$ & 50230.738543 & 50230.738543 & 50230.738567 & -0.74 & 4.266364 \\ 
$6s28s \, {}^1 \! S_0$ & 50248.236248 & 50248.236248 & 50248.236241 & 0.22 & 4.267477 \\ 
$6s29s \, {}^1 \! S_0$ & 50263.659051 & 50263.659051 & 50263.659169 & -3.56 & 4.268432 \\ 
$6s30s \, {}^1 \! S_0$ & 50277.322237 & 50277.322237 & 50277.322373 & -4.10 & 4.269260 \\ 
$6s31s \, {}^1 \! S_0$ & 50289.483450 & 50289.483450 & 50289.483399 & 1.53 & 4.269983 \\ 
$6s32s \, {}^1 \! S_0$ & 50300.354170 & 50300.354170 & 50300.354421 & -7.51 & 4.270620 \\ 
$6s33s \, {}^1 \! S_0$ & 50310.111320 & 50310.111320 & 50310.111484 & -4.92 & 4.271184 \\ 
$6s34s \, {}^1 \! S_0$ & 50318.901535 & 50318.901648$^{\bullet}$ & 50318.901631 & 0.49 & 4.271687 \\ 
$6s35s \, {}^1 \! S_0$ & 50326.848499 & 50326.848456$^{\bullet}$ & 50326.848444 & 0.37 & 4.272137 \\ 
$6s36s \, {}^1 \! S_0$ & 50334.056419 & 50334.056400$^{\bullet}$ & 50334.056387 & 0.39 & 4.272542 \\ 
$6s37s \, {}^1 \! S_0$ & 50340.614223 & 50340.614269$^{\bullet}$ & 50340.614251 & 0.54 & 4.272907 \\ 
$6s38s \, {}^1 \! S_0$ & 50346.597829 & 50346.597899$^{\bullet}$ & 50346.597886 & 0.38 & 4.273238 \\ 
$6s39s \, {}^1 \! S_0$ & 50352.072416 & 50352.072403$^{\bullet}$ & 50352.072402 & 0.04 & 4.273539 \\ 
$6s40s \, {}^1 \! S_0$ & 50357.093890 & 50357.093898$^{\bullet}$ & 50357.093937 & -1.18 & 4.273814 \\ 
$6s41s \, {}^1 \! S_0$ & 50361.711017 & 50361.711104$^{\bullet}$ & 50361.711106 & -0.07 & 4.274066 \\ 
$6s42s \, {}^1 \! S_0$ & 50365.966161 & 50365.966161 & 50365.966175 & -0.42 & 4.274297 \\ 
$6s43s \, {}^1 \! S_0$ & 50369.896147 & 50369.896018$^{\bullet}$ & 50369.896029 & -0.34 & 4.274509 \\ 
$6s44s \, {}^1 \! S_0$ & 50373.532996 & 50373.532962$^{\bullet}$ & 50373.532977 & -0.45 & 4.274705 \\ 
$6s45s \, {}^1 \! S_0$ & 50376.905462 & 50376.905396$^{\bullet}$ & 50376.905413 & -0.51 & 4.274886 \\ 
$6s46s \, {}^1 \! S_0$ & 50380.038430 & 50380.038367$^{\bullet}$ & 50380.038373 & -0.19 & 4.275054 \\ 
$6s47s \, {}^1 \! S_0$ & 50382.953980 & 50382.954005$^{\bullet}$ & 50382.954002 & 0.09 & 4.275209 \\ 
$6s48s \, {}^1 \! S_0$ & 50385.671794 & 50385.671949$^{\bullet}$ & 50385.671941 & 0.23 & 4.275354 \\ 
$6s49s \, {}^1 \! S_0$ & 50388.209683 & 50388.209676$^{\bullet}$ & 50388.209665 & 0.34 & 4.275489 \\ 
$6s50s \, {}^1 \! S_0$ & 50390.582791 & 50390.582772$^{\bullet}$ & 50390.582757 & 0.45 & 4.275615 \\ 
$6s51s \, {}^1 \! S_0$ & 50392.805062 & 50392.805174$^{\bullet}$ & 50392.805154 & 0.59 & 4.275732 \\ 
$6s52s \, {}^1 \! S_0$ & 50394.889104 & 50394.889374$^{\bullet}$ & 50394.889349 & 0.74 & 4.275842 \\ 
$6s53s \, {}^1 \! S_0$ & 50396.846524 & 50396.846583$^{\bullet}$ & 50396.846566 & 0.51 & 4.275946 \\ 
$6s54s \, {}^1 \! S_0$ & 50398.686864 & 50398.686864 & 50398.686912 & -1.46 & 4.276043 \\ 
$6s55s \, {}^1 \! S_0$ & 50400.419329 & 50400.419329 & 50400.419511 & -5.45 & 4.276134 \\ 
$6s56s \, {}^1 \! S_0$ & 50402.052459 & 50402.052459 & 50402.052610 & -4.52 & 4.276219 \\ 
$6s57s \, {}^1 \! S_0$ & 50403.593725 & 50403.593725 & 50403.593684 & 1.23 & 4.276300 \\ 
$6s58s \, {}^1 \! S_0$ & 50405.049399 & 50405.049399 & 50405.049519 & -3.60 & 4.276377 \\ 
$6s59s \, {}^1 \! S_0$ & 50406.426285 & 50406.426285 & 50406.426287 & -0.07 & 4.276449 \\ 
$6s60s \, {}^1 \! S_0$ & 50407.729587 & 50407.729587 & 50407.729611 & -0.72 & 4.276517 \\ 
$6s61s \, {}^1 \! S_0$ & 50408.964441 & 50408.964441 & 50408.964623 & -5.46 & 4.276581 \\ 
$6s62s \, {}^1 \! S_0$ & 50410.135851 & 50410.135851 & 50410.136014 & -4.90 & 4.276642 \\ 
$6s63s \, {}^1 \! S_0$ & 50411.248021 & 50411.248021 & 50411.248081 & -1.79 & 4.276700 \\ 
$6s64s \, {}^1 \! S_0$ & 50412.304819 & 50412.304819 & 50412.304761 & 1.72 & 4.276755 \\ 
$6s65s \, {}^1 \! S_0$ & 50413.309647 & 50413.309647 & 50413.309674 & -0.81 & 4.276808 \\ 
$6s66s \, {}^1 \! S_0$ & 50414.266242 & 50414.266242 & 50414.266145 & 2.90 & 4.276857 \\ 
$6s67s \, {}^1 \! S_0$ & 50415.177206 & 50415.177206 & 50415.177239 & -0.99 & 4.276905 \\ 
$6s68s \, {}^1 \! S_0$ & 50416.045807 & 50416.045807 & 50416.045781 & 0.77 & 4.276950 \\ 
$6s69s \, {}^1 \! S_0$ & 50416.874380 & 50416.874380 & 50416.874380 & -0.02 & 4.276993 \\ 
$6s70s \, {}^1 \! S_0$ & 50417.665460 & 50417.665460 & 50417.665449 & 0.33 & 4.277034 \\ 
$6s71s \, {}^1 \! S_0$ & 50418.421250 & 50418.421250 & 50418.421219 & 0.91 & 4.277073 \\ 
$6s72s \, {}^1 \! S_0$ & 50419.143750 & 50419.143750 & 50419.143761 & -0.34 & 4.277111 \\ 
$6s73s \, {}^1 \! S_0$ & 50419.835028 & 50419.835028 & 50419.834993 & 1.03 & 4.277146 \\ 
$6s74s \, {}^1 \! S_0$ & 50420.496619 & 50420.496619 & 50420.496700 & -2.42 & 4.277181 \\ 
$6s75s \, {}^1 \! S_0$ & 50421.130458 & 50421.130458 & 50421.130538 & -2.39 & 4.277213 \\ 
$6s76s \, {}^1 \! S_0$ & 50421.737944 & 50421.737944 & 50421.738051 & -3.20 & 4.277245 \\ 
$6s77s \, {}^1 \! S_0$ & 50422.320814 & 50422.320814 & 50422.320676 & 4.12 & 4.277275 \\ 
$6s78s \, {}^1 \! S_0$ & 50422.879734 & 50422.879734 & 50422.879756 & -0.66 & 4.277304 \\ 
$6s79s \, {}^1 \! S_0$ & 50423.416439 & 50423.416439 & 50423.416541 & -3.05 & 4.277332 \\ 
$6s80s \, {}^1 \! S_0$ & 50423.932196 & 50423.932196 & 50423.932201 & -0.16 & 4.277358 \\ 
\end{longtable*}

\begin{longtable*}{@{\extracolsep{\fill}}cccccc}
\caption{Experimental and theoretical energies of the $6snd \,{}^{3,1}\!D_2$ Rydberg series of $^{174}$Yb. $E_{LSp}$ is the new laser spectroscopy data. $E_{Cb}$ is the chosen combination of experimental data to be analyzed with MQDT: Data followed with $^{\blacktriangle}$ is taken from \cite{Aymar_1980}, data followed with $^{\bullet}$ is extrapolated from $6s(n+1)s$ with the microwave data presented in Table \ref{6snsnd_MW}, data without sign is from $E_{LSp}$. $E_{Th}$ is the theoretical energy obtained with our MQDT model presented in Table \ref{6snd_MQDT}. The next column corresponds to the difference between theoretical and experimental energies expressed in MHz and the last column is the quantum defect inferred from the MQDT model, computed assuming the ionization limit at 50443.07040 cm$^{-1}$.} \label{6snd_values} \\

\hline \hline  \\[-2.0ex] Assignment & $E_{new}$ (cm$^{-1}$) & $E_{comb.}$ (cm$^{-1}$) & $E_{th}$ (cm$^{-1}$) & $E_{exp}-E_{th}$ (MHz) & Q. Defect \\ \hline \\[-2.0ex]
\endfirsthead

\multicolumn{6}{c}%
{{\bfseries \tablename\ \thetable{} -- continued from previous page}} \\
\hline  \\[-2.0ex] \multicolumn{1}{c}{Assignment} &
\multicolumn{1}{c}{$E_{new}$ (cm$^{-1}$)} &
\multicolumn{1}{c}{$E_{comb.}$ (cm$^{-1}$)} &
\multicolumn{1}{c}{$E_{th}$ (cm$^{-1}$)} &
\multicolumn{1}{c}{$E_{exp}-E_{th}$ (MHz)} &
\multicolumn{1}{c}{Q. Defect} \\ \hline  \\[-2.0ex]
\endhead

\hline \multicolumn{6}{|r|}{{\rule{0pt}{10pt} Continued on next page}} \\ \hline
\endfoot

\hline \hline
\endlastfoot

$6s8d \,{}^1\!D_2$ & $\,$ & 46405.62$^{\blacktriangle}$ & 46403.220405 & 71940 & 2.788127 \\ 
$6s8d \,{}^3\!D_2$ & $\,$ & 46467.69$^{\blacktriangle}$ & 46468.006828 & -9499 & 2.745826 \\ 
$6p^2 \,{}^1\!D_2$ & $\,$ & 47420.97$^{\blacktriangle}$ & 47421.203636 & -7005 & - \\ 
$6s9d \,{}^1\!D_2$ & $\,$ & 47821.74$^{\blacktriangle}$ & 47822.561106 & -24620 & 2.528818 \\ 
$6s9d \,{}^3\!D_2$ & $\,$ & 47634.40$^{\blacktriangle}$ & 47634.215002 & 5546 & 2.749543 \\ 
$6s10d \,{}^1\!D_2$ & $\,$ & 48403.49$^{\blacktriangle}$ & 48394.839809 & 259300 & 2.680405 \\ 
$6s10d \,{}^3\!D_2$ & $\,$ & 48357.63$^{\blacktriangle}$ & 48353.122376 & 135100 & 2.753826 \\ 
$4f^{13}5d6s6p \,B$ & $\,$ & 48762.52$^{\blacktriangle}$ & 48764.627471 & -63180 & - \\ 
$6s11d \,{}^1\!D_2$ & $\,$ & 48883.12$^{\blacktriangle}$ & 48876.949819 & 185000 & 2.629259 \\ 
$6s11d \,{}^3\!D_2$ & $\,$ & 48838.14$^{\blacktriangle}$ & 48836.053145 & 62560 & 2.736459 \\ 
$6s12d \,{}^1\!D_2$ & $\,$ & 49176000$^{\blacktriangle}$ & 49177.269330 & -38050 & 2.689055 \\ 
$6s12d \,{}^3\!D_2$ & $\,$ & 49161.12$^{\blacktriangle}$ & 49161.794071 & -20210 & 2.745455 \\ 
$6s13d \,{}^1\!D_2$ & $\,$ & 49408.58$^{\blacktriangle}$ & 49409.033108 & -13580 & 2.698312 \\ 
$6s13d \,{}^3\!D_2$ & $\,$ & 49399.10$^{\blacktriangle}$ & 49399.204959 & -3147 & 2.746923 \\ 
$6s14d \,{}^1\!D_2$ & $\,$ & 49583.28$^{\blacktriangle}$ & 49583.268280 & 351 & 2.702633 \\ 
$6s14d \,{}^3\!D_2$ & $\,$ & 49576.36$^{\blacktriangle}$ & 49576.395882 & -1076 & 2.747514 \\ 
$6s15d \,{}^1\!D_2$ & $\,$ & 49717.15$^{\blacktriangle}$ & 49717.113377 & 1097 & 2.705213 \\ 
$6s15d \,{}^3\!D_2$ & $\,$ & 49712.11$^{\blacktriangle}$ & 49712.054327 & 1669 & 2.747830 \\ 
$6s16d \,{}^1\!D_2$ & $\,$ & 49822.08$^{\blacktriangle}$ & 49822.053773 & 786 & 2.706943 \\ 
$6s16d \,{}^3\!D_2$ & $\,$ & 49818.19$^{\blacktriangle}$ & 49818.197542 & -227 & 2.748024 \\ 
$6s17d \,{}^1\!D_2$ & $\,$ & 49905.79$^{\blacktriangle}$ & 49905.817505 & -825 & 2.708195 \\ 
$6s17d \,{}^3\!D_2$ & $\,$ & 49902.78$^{\blacktriangle}$ & 49902.800583 & -618 & 2.748154 \\ 
$6s18d \,{}^1\!D_2$ & $\,$ & 49973.69$^{\blacktriangle}$ & 49973.727869 & -1136 & 2.709154 \\ 
$6s18d \,{}^3\!D_2$ & $\,$ & 49971.34$^{\blacktriangle}$ & 49971.318753 & 636 & 2.748247 \\ 
$6s19d \,{}^1\!D_2$ & $\,$ & 50029.61$^{\blacktriangle}$ & 50029.540423 & 2085 & 2.709929 \\ 
$6s19d \,{}^3\!D_2$ & $\,$ & 50027.66$^{\blacktriangle}$ & 50027.584510 & 2263 & 2.748317 \\ 
$6s20d \,{}^1\!D_2$ & $\,$ & 50076.05$^{\blacktriangle}$ & 50075.963353 & 2597 & 2.710592 \\ 
$6s20d \,{}^3\!D_2$ & $\,$ & 50074.44$^{\blacktriangle}$ & 50074.353678 & 2587 & 2.748373 \\ 
$6s21d \,{}^1\!D_2$ & $\,$ & 50115.18$^{\blacktriangle}$ & 50114.988242 & 5748 & 2.711202 \\ 
$6s21d \,{}^3\!D_2$ & $\,$ & 50113.67$^{\blacktriangle}$ & 50113.648840 & 634 & 2.748420 \\ 
$6s22d \,{}^1\!D_2$ & $\,$ & 50148.59$^{\blacktriangle}$ & 50148.105429 & 1.453e+004 & 2.711820 \\ 
$6s22d \,{}^3\!D_2$ & $\,$ & 50147.03$^{\blacktriangle}$ & 50146.981496 & 1454 & 2.748463 \\ 
$6s23d \,{}^1\!D_2$ & 50176.447184 & 50176.447184 & 50176.447238 & -1.63 & 2.712546 \\ 
$6s23d \,{}^3\!D_2$ & 50175.499462 & 50175.499462 & 50175.499494 & -0.98 & 2.748508 \\ 
$6s24d \,{}^1\!D_2$ & 50200.884090 & 50200.884090 & 50200.884131 & -1.23 & 2.713624 \\ 
$6s24d \,{}^3\!D_2$ & 50200.087072 & 50200.087072 & 50200.087055 & 0.50 & 2.748567 \\ 
$6s25d \,{}^1\!D_2$ & 50222.082355 & 50222.082355 & 50222.082247 & 3.23 & 2.716060 \\ 
$6s25d \,{}^3\!D_2$ & 50221.433573 & 50221.433573 & 50221.433546 & 0.79 & 2.748695 \\ 
$6s26d \,{}^1\!D_2$ & 50240.290618 & 50240.290618 & 50240.290626 & -0.24 & 2.737084 \\ 
$6s26d \,{}^3\!D_2$ & 50240.035775 & 50240.035775 & 50240.035785 & -0.30 & 2.751688 \\ 
$4f^{13}5d6s6p \,A$ & 50244.240350 & 50244.240350 & 50244.240348 & 0 & - \\ 
$6s27d \,{}^1\!D_2$ & 50257.162357 & 50257.162357 & 50257.162284 & 2.18 & 2.704418 \\ 
$6s27d \,{}^3\!D_2$ & 50256.489825 & 50256.489825 & 50256.489877 & -1.56 & 2.748236 \\ 
$6s28d \,{}^1\!D_2$ & 50271.516353 & 50271.516211$^{\bullet}$ & 50271.516216 & -0.16 & 2.708428 \\ 
$6s28d \,{}^3\!D_2$ & 50270.973044 & 50270.973044 & 50270.972943 & 3.02 & 2.748380 \\ 
$6s29d \,{}^1\!D_2$ & 50284.302933 & 50284.303041$^{\bullet}$ & 50284.303034 & 0.19 & 2.709681 \\ 
$6s29d \,{}^3\!D_2$ & 50283.833875 & 50283.833875 & 50283.833968 & -2.81 & 2.748431 \\ 
$6s30d \,{}^1\!D_2$ & 50295.718230 & 50295.718328$^{\bullet}$ & 50295.718332 & -0.12 & 2.710325 \\ 
$6s30d \,{}^3\!D_2$ & 50295.305744 & 50295.305744 & 50295.305642 & 3.07 & 2.748460 \\ 
$6s31d \,{}^1\!D_2$ & 50305.947573 & 50305.947641$^{\bullet}$ & 50305.947632 & 0.26 & 2.710733 \\ 
$6s31d \,{}^3\!D_2$ & 50305.580853 & 50305.580853 & 50305.580969 & -3.47 & 2.748480 \\ 
$6s32d \,{}^1\!D_2$ & 50315.148605 & 50315.148609$^{\bullet}$ & 50315.148611 & -0.07 & 2.711022 \\ 
$6s32d \,{}^3\!D_2$ & 50314.820778 & 50314.820778 & 50314.820656 & 3.67 & 2.748494 \\ 
$6s33d \,{}^1\!D_2$ & 50323.454151 & 50323.454188$^{\bullet}$ & 50323.454197 & -0.28 & 2.711242 \\ 
$6s33d \,{}^3\!D_2$ & 50323.159280 & 50323.159280 & 50323.159324 & -1.34 & 2.748506 \\ 
$6s34d \,{}^1\!D_2$ & 50330.976689 & 50330.976713$^{\bullet}$ & 50330.976718 & -0.16 & 2.711416 \\ 
$6s34d \,{}^3\!D_2$ & 50330.710371 & 50330.710371 & 50330.710422 & -1.53 & 2.748515 \\ 
$6s35d \,{}^1\!D_2$ & 50337.811617 & 50337.811520$^{\bullet}$ & 50337.811506 & 0.42 & 2.711559 \\ 
$6s35d \,{}^3\!D_2$ & 50337.570183 & 50337.570183 & 50337.570088 & 2.84 & 2.748523 \\ 
$6s36d \,{}^1\!D_2$ & 50344.039792 & 50344.039856$^{\bullet}$ & 50344.039852 & 0.12 & 2.711679 \\ 
$6s36d \,{}^3\!D_2$ & 50343.820307 & 50343.820307 & 50343.820231 & 2.27 & 2.748530 \\ 
$6s37d \,{}^1\!D_2$ & 50349.731397 & 50349.731398$^{\bullet}$ & 50349.731403 & -0.14 & 2.711782 \\ 
$6s37d \,{}^3\!D_2$ & 50349.530858 & 50349.530858 & 50349.530980 & -3.67 & 2.748536 \\ 
$6s38d \,{}^1\!D_2$ & 50354.946004 & 50354.946093$^{\bullet}$ & 50354.946095 & -0.07 & 2.711872 \\ 
$6s38d \,{}^3\!D_2$ & 50354.762544 & 50354.762544 & 50354.762662 & -3.54 & 2.748541 \\ 
$6s39d \,{}^1\!D_2$ & 50359.735651 & 50359.735749$^{\bullet}$ & 50359.735736 & 0.40 & 2.711950 \\ 
$6s39d \,{}^3\!D_2$ & 50359.567535 & 50359.567535 & 50359.567399 & 4.06 & 2.748545 \\ 
$6s40d \,{}^1\!D_2$ & 50364.145301 & 50364.145263$^{\bullet}$ & 50364.145285 & -0.65 & 2.712019 \\ 
$6s40d \,{}^3\!D_2$ & 50363.990461 & 50363.990461 & 50363.990415 & 1.39 & 2.748550 \\ 
$6s41d \,{}^1\!D_2$ & 50368.213916 & 50368.213919$^{\bullet}$ & 50368.213914 & 0.15 & 2.712081 \\ 
$6s41d \,{}^3\!D_2$ & 50368.071084 & 50368.071084 & 50368.071098 & -0.42 & 2.748553 \\ 
$6s42d \,{}^1\!D_2$ & 50371.975785 & 50371.975864$^{\bullet}$ & 50371.975878 & -0.42 & 2.712137 \\ 
$6s42d \,{}^3\!D_2$ & 50371.843894 & 50371.843894 & 50371.843886 & 0.24 & 2.748556 \\ 
$6s43d \,{}^1\!D_2$ & 50375.461130 & 50375.461220$^{\bullet}$ & 50375.461235 & -0.45 & 2.712187 \\ 
$6s43d \,{}^3\!D_2$ & 50375.338979 & 50375.338979 & 50375.338992 & -0.41 & 2.748559 \\ 
$6s44d \,{}^1\!D_2$ & 50378.696435 & 50378.696447$^{\bullet}$ & 50378.696450 & -0.08 & 2.712233 \\ 
$6s44d \,{}^3\!D_2$ & 50378.583090 & 50378.583090 & 50378.583014 & 2.29 & 2.748562 \\ 
$6s45d \,{}^1\!D_2$ & 50381.704916 & 50381.704915$^{\bullet}$ & 50381.704897 & 0.54 & 2.712274 \\ 
$6s45d \,{}^3\!D_2$ & 50381.599376 & 50381.599376 & 50381.599436 & -1.81 & 2.748565 \\ 
$6s46d \,{}^1\!D_2$ & 50384.507121 & 50384.507300$^{\bullet}$ & 50384.507283 & 0.51 & 2.712312 \\ 
$6s46d \,{}^3\!D_2$ & 50384.409120 & 50384.409120 & 50384.409063 & 1.69 & 2.748567 \\ 
$6s47d \,{}^1\!D_2$ & 50387.121930 & 50387.122020$^{\bullet}$ & 50387.122005 & 0.45 & 2.712347 \\ 
$6s47d \,{}^3\!D_2$ & 50387.030400 & 50387.030400 & 50387.030375 & 0.74 & 2.748569 \\ 
$6s48d \,{}^1\!D_2$ & 50389.565354 & 50389.565438$^{\bullet}$ & 50389.565451 & -0.40 & 2.712380 \\ 
$6s48d \,{}^3\!D_2$ & 50389.479828 & 50389.479828 & 50389.479832 & -0.12 & 2.748571 \\ 
$6s49d \,{}^1\!D_2$ & 50391.852203 & 50391.852203 & 50391.852260 & -1.70 & 2.712409 \\ 
$6s49d \,{}^3\!D_2$ & 50391.772081 & 50391.772081 & 50391.772135 & -1.61 & 2.748573 \\ 
$6s50d \,{}^1\!D_2$ & 50393.995485 & 50393.995485 & 50393.995538 & -1.58 & 2.712437 \\ 
$6s50d \,{}^3\!D_2$ & 50393.920367 & 50393.920367 & 50393.920445 & -2.33 & 2.748575 \\ 
$6s51d \,{}^1\!D_2$ & 50396.006944 & 50396.006944 & 50396.007050 & -3.17 & 2.712463 \\ 
$6s51d \,{}^3\!D_2$ & 50395.936562 & 50395.936562 & 50395.936575 & -0.38 & 2.748576 \\ 
$6s52d \,{}^1\!D_2$ & 50397.897318 & 50397.897318 & 50397.897379 & -1.83 & 2.712486 \\ 
$6s52d \,{}^3\!D_2$ & 50397.831139 & 50397.831139 & 50397.831150 & -0.32 & 2.748578 \\ 
$6s53d \,{}^1\!D_2$ & 50399.676015 & 50399.676015 & 50399.676067 & -1.57 & 2.712509 \\ 
$6s53d \,{}^3\!D_2$ & 50399.613772 & 50399.613772 & 50399.613748 & 0.70 & 2.748579 \\ 
$6s54d \,{}^1\!D_2$ & 50401.351708 & 50401.351708 & 50401.351736 & -0.84 & 2.712529 \\ 
$6s54d \,{}^3\!D_2$ & 50401.292934 & 50401.292934 & 50401.293024 & -2.72 & 2.748580 \\ 
$6s55d \,{}^1\!D_2$ & 50402.932001 & 50402.932001 & 50402.932189 & -5.64 & 2.712549 \\ 
$6s55d \,{}^3\!D_2$ & 50402.876696 & 50402.876696 & 50402.876811 & -3.46 & 2.748581 \\ 
$6s56d \,{}^1\!D_2$ & 50404.424367 & 50404.424367 & 50404.424506 & -4.17 & 2.712567 \\ 
$6s56d \,{}^3\!D_2$ & 50404.372197 & 50404.372197 & 50404.372214 & -0.51 & 2.748583 \\ 
$6s57d \,{}^1\!D_2$ & 50405.835009 & 50405.835009 & 50405.835120 & -3.34 & 2.712584 \\ 
$6s57d \,{}^3\!D_2$ & 50405.785642 & 50405.785642 & 50405.785688 & -1.38 & 2.748584 \\ 
$6s58d \,{}^1\!D_2$ & 50407.169799 & 50407.169799 & 50407.169889 & -2.71 & 2.712600 \\ 
$6s58d \,{}^3\!D_2$ & 50407.123100 & 50407.123100 & 50407.123111 & -0.35 & 2.748585 \\ 
$6s59d \,{}^1\!D_2$ & 50408.434007 & 50408.434007 & 50408.434154 & -4.40 & 2.712616 \\ 
$6s59d \,{}^3\!D_2$ & 50408.389843 & 50408.389843 & 50408.389843 & -0.01 & 2.748585 \\ 
$6s60d \,{}^1\!D_2$ & 50409.632637 & 50409.632637 & 50409.632793 & -4.69 & 2.712630 \\ 
$6s60d \,{}^3\!D_2$ & 50409.590674 & 50409.590674 & 50409.590779 & -3.17 & 2.748586 \\ 
$6s61d \,{}^1\!D_2$ & 50410.770157 & 50410.770157 & 50410.770273 & -3.47 & 2.712643 \\ 
$6s61d \,{}^3\!D_2$ & 50410.730329 & 50410.730329 & 50410.730399 & -2.10 & 2.748587 \\ 
$6s62d \,{}^1\!D_2$ & 50411.850638 & 50411.850638 & 50411.850683 & -1.34 & 2.712656 \\ 
$6s62d \,{}^3\!D_2$ & 50411.812745 & 50411.812745 & 50411.812806 & -1.83 & 2.748588 \\ 
$6s63d \,{}^1\!D_2$ & 50412.877748 & 50412.877748 & 50412.877778 & -0.91 & 2.712668 \\ 
$6s63d \,{}^3\!D_2$ & 50412.841590 & 50412.841590 & 50412.841767 & -5.31 & 2.748589 \\ 
$6s64d \,{}^1\!D_2$ & 50413.854958 & 50413.854958 & 50413.855010 & -1.56 & 2.712680 \\ 
$6s64d \,{}^3\!D_2$ & 50413.820734 & 50413.820734 & 50413.820744 & -0.30 & 2.748589 \\ 
$6s65d \,{}^1\!D_2$ & 50414.785468 & 50414.785468 & 50414.785554 & -2.60 & 2.712690 \\ 
$6s65d \,{}^3\!D_2$ & 50414.752979 & 50414.752979 & 50414.752922 & 1.70 & 2.748590 \\ 
$6s66d \,{}^1\!D_2$ & 50415.672415 & 50415.672415 & 50415.672339 & 2.28 & 2.712701 \\ 
$6s66d \,{}^3\!D_2$ & 50415.641260 & 50415.641260 & 50415.641238 & 0.66 & 2.748591 \\ 
$6s67d \,{}^1\!D_2$ & 50416.518067 & 50416.518067 & 50416.518065 & 0.07 & 2.712711 \\ 
$6s67d \,{}^3\!D_2$ & 50416.488379 & 50416.488379 & 50416.488400 & -0.64 & 2.748591 \\ 
$6s68d \,{}^1\!D_2$ & 50417.325292 & 50417.325292 & 50417.325228 & 1.92 & 2.712720 \\ 
$6s68d \,{}^3\!D_2$ & 50417.296939 & 50417.296939 & 50417.296913 & 0.77 & 2.748592 \\ 
$6s69d \,{}^1\!D_2$ & 50418.096092 & 50418.096092 & 50418.096137 & -1.37 & 2.712729 \\ 
$6s69d \,{}^3\!D_2$ & 50418.069206 & 50418.069206 & 50418.069092 & 3.42 & 2.748592 \\ 
$6s70d \,{}^1\!D_2$ & 50418.833001 & 50418.833001 & 50418.832933 & 2.04 & 2.712737 \\ 
$6s70d \,{}^3\!D_2$ & 50418.807117 & 50418.807117 & 50418.807081 & 1.07 & 2.748593 \\ 
$6s71d \,{}^1\!D_2$ & 50419.537689 & 50419.537689 & 50419.537598 & 2.74 & 2.712745 \\ 
$6s71d \,{}^3\!D_2$ & 50419.512872 & 50419.512872 & 50419.512870 & 0.04 & 2.748593 \\ 
$6s72d \,{}^1\!D_2$ & 50420.212022 & 50420.212022 & 50420.211973 & 1.47 & 2.712753 \\ 
$6s72d \,{}^3\!D_2$ & 50420.188339 & 50420.188339 & 50420.188306 & 0.98 & 2.748594 \\ 
$6s73d \,{}^1\!D_2$ & 50420.857736 & 50420.857736 & 50420.857770 & -1.03 & 2.712760 \\ 
$6s73d \,{}^3\!D_2$ & 50420.835120 & 50420.835120 & 50420.835104 & 0.48 & 2.748594 \\ 
$6s74d \,{}^1\!D_2$ & 50421.476497 & 50421.476497 & 50421.476582 & -2.54 & 2.712767 \\ 
$6s74d \,{}^3\!D_2$ & 50421.455115 & 50421.455115 & 50421.454860 & 7.64 & 2.748595 \\ 
$6s75d \,{}^1\!D_2$ & 50422.069907 & 50422.069907 & 50422.069890 & 0.50 & 2.712774 \\ 
$6s75d \,{}^3\!D_2$ & 50422.049160 & 50422.049160 & 50422.049062 & 2.95 & 2.748595 \\ 
$6s76d \,{}^1\!D_2$ & 50422.639235 & 50422.639235 & 50422.639078 & 4.70 & 2.712781 \\ 
$6s76d \,{}^3\!D_2$ & 50422.619154 & 50422.619154 & 50422.619094 & 1.79 & 2.748596 \\ 
$6s77d \,{}^1\!D_2$ & 50423.185412 & 50423.185412 & 50423.185435 & -0.71 & 2.712787 \\ 
$6s77d \,{}^3\!D_2$ & 50423.166266 & 50423.166266 & 50423.166251 & 0.45 & 2.748596 \\ 
$6s78d \,{}^1\!D_2$ & 50423.710176 & 50423.710176 & 50423.710167 & 0.27 & 2.712793 \\ 
$6s78d \,{}^3\!D_2$ & 50423.691829 & 50423.691829 & 50423.691740 & 2.67 & 2.748596 \\ 
$6s79d \,{}^1\!D_2$ & 50424.214458 & 50424.214458 & 50424.214399 & 1.76 & 2.712798 \\ 
$6s79d \,{}^3\!D_2$ & 50424.196846 & 50424.196846 & 50424.196690 & 4.67 & 2.748597 \\ 
$6s80d \,{}^1\!D_2$ & 50424.699260 & 50424.699260 & 50424.699186 & 2.21 & 2.712804 \\ 
$6s80d \,{}^3\!D_2$ & 50424.682115 & 50424.682115 & 50424.682158 & -1.30 & 2.748597 \\ 
\end{longtable*}

\bibliographystyle{apsrev4-1}

\begin{thebibliography}{40}

\bibitem{Aymar_1980} Aymar, M., Debarre, A., \& Robaux, O.\ 1980, Journal of Physics B Atomic Molecular Physics, 13, 1089

\bibitem{Ovsiannikov_2011} Ovsiannikov, V.~D., Derevianko, A., \& Gibble, K.\ 2011, Physical Review Letters, 107, 093003

\bibitem{Bowden_2017} Bowden, W., Hobson, R., Huillery, P., Gill, P., Jones, M.~P.~A., \& Hill, I.~R.\ 2017, \pra, 96, 023419

\bibitem{Safronova_2012} Safronova, M.~S., Porsev, S.~G., Kozlov, M.~G., \& Clark, C.~W. \ 2012, \pra, 85, 052506

\bibitem{Tamm_2016} Huntemann, N., Sanner, C., Lipphardt, B., Tamm, Ch., and Peik, E.\ 2016, \prl, 116, 063001

\bibitem{Jiang_2009}
Jiang, D., Arora, B., Safronova, M.~S., \& Clark, C.~W.\ 2009, Journal of Physics B Atomic Molecular Physics, 42, 154020

\bibitem{Ward_1996} Ward Jr., R.~F., Sturrus, W.~G., \& Lundeen, S.~R. \ 1996, \pra, 53, 113

\bibitem{Weimer_2010}
Weimer, H., M{\"u}ller, M., Lesanovsky, I., Zoller, P., B{\"u}chler, H.~P.\ 2010, Nature Physics, 6, 382

\bibitem{Saffman_2010}
Saffman, M., Walker, T.~G., \& M{\o}lmer, K.\ 2010, Reviews of Modern Physics, 82, 2313

\bibitem{Lukin_2001}
Lukin, M.~D., Fleischhauer, M., Cote, R., Duan, L.-M., Jaksch, D., Cirac, J.~I., \& Zoller, P.\ 2001, \prl, 87, 037901

\bibitem{Urban_2009}
Urban, E., Johnson, T.~A., Henage, T., et al.\ 2009, Nature Physics, 5, 110

\bibitem{Gaetan_2009}
Ga{\"e}tan, A., Miroshnychenko, Y., Wilk, T., et al.\ 2009, Nature Physics, 5, 115

\bibitem{Dudin_2012} Y.~O. Dudin, A. Kuzmich, \ 2012, Science, 336, 887

\bibitem{Bell_1991} Bell, A.~S., Gill, P., Klein, H.~A., Levick, A.~P., Tamm, C., \& Schnier, D. \ 1991, \pra, 44, R20

\bibitem{Cooke_1978}
Cooke, W.~E., Gallagher, T.~F., Edelstein, S.~A.,
\& Hill, R.~M.\ 1978, Physical Review Letters, 40, 178

\bibitem{McQuillen_2013} McQuillen, P., Zhang, X., Strickler, T., Dunning, F.~B., \& Killian, T.~C. \ 2013, \pra, 87, 013407

\bibitem{Lochead_2013} Lochead, G., Boddy, D., Sadler, D.~P., Adams, C.~S., \& Jones, M.~P.~A. \ 2013, \pra, 87, 053409

\bibitem{Schauss_2012} Schau{\ss}, P., Cheneau, M., Endres, M., et al.\ 2012, \nat, 491, 87

\bibitem{Gunter_2012} G{\"u}nter, G., Robert-de-Saint-Vincent, M., Schempp, H., et al.\ 2012, Physical Review Letters, 108, 013002

\bibitem{Dutta_2000} Dutta, S.~K., Guest, J.~R., Feldbaum, D., Walz\-Flannigan, A., \& Raithel, G. \ 2000, Physical Review Letters, 85, 5551

\bibitem{Anderson_2011} Anderson, S.~E., Younge, K.~C., \& Raithel, G. \ 2000, Physical Review Letters, 107, 263001

\bibitem{Saffman_2005} Saffman, M., \& Walker, T.~G. \ 2005, \pra, 72, 022347

\bibitem{Wyart_1979} Wyart, J.-F., \& Camus, P.\ 1979, \physscr, 20, 43

\bibitem{Camus_1980}
Camus, P., Debarre, A., \& Morillon, C.\ 1980, Journal of Physics B Atomic Molecular Physics, 13, 1073

\bibitem{Xu_1994} Xu, C.~B., Xu, X.~Y., Huang, W., Xue, M., \& Chen, D.~Y.\ 1994, Journal of Physics B Atomic Molecular Physics, 27, 3905

\bibitem{Aymar_1984} Aymar, M., Champeau, R.~J., Delsart, C., \& Robaux, O.\ 1984, Journal of Physics B Atomic Molecular Physics, 17, 3645

\bibitem{Maeda_1992} Maeda, H., Matsuo,  Y., Takami, M., Suzuki, A. \ 1992, \pra, 45, 1732

\bibitem{Shuman_2007} Shuman, E. S., Nunkaew, J., and Gallagher, T. F.  \ 2007, \pra, 75, 044501


\bibitem{Kuwamoto_1999} Kuwamoto, T., Honda, K., Takahashi, Y., \& Yabuzaki, T.\ 1999, \pra, 60, R745

\bibitem{Cooke_1979} Cooke, W.~E., \& Gallagher, T.~F.\ 1979, \pra, 19, 2151

\bibitem{Kleinert_2016} Kleinert, M., Gold Dahl, M.~E., \& Bergeson, S.\ 2016, \pra, 94, 052511

\bibitem{Fano_1975} Fano, U. \ 1975, J. Opt. Soc. Am., 65, 979

\bibitem{Seaton_1966} Seaton, M.~J. \ 1966, Proc. Phys. Soc. (London), 88, 801

\bibitem{Cooke_1985} Cooke, W.~E., \& Cromer, C.~L.\ 1985, \pra, 32, 2725

\bibitem{Aymar_1996} Aymar, M., Greene, C.~H., \& Luc-Koenig, E. \ 1996, Rev.~Mod.~Phys., 68, 1015

\bibitem{Vaillant_2014} Vaillant, C.~L., Jones, M.~P.~A., \& Potvliege, R.~M. \ 2014, Journal of Physics B Atomic Molecular Physics, 47, 155001

\bibitem{Lu_1970} Lu, K.~T., \& Fano, U.\ 1970, \pra, 2, 81

\bibitem{Seaton_1983} Seaton, M. J., \ 1983, Rep. Prog. Phys. 46, 167

\bibitem{Lecomte_1987} Lecomte, J.~M. \ 1987, Journal of Physics B Atomic Molecular Physics, 20, 3645


\bibitem{Osterwalder_2004} Osterwalder, A., W\"uest, A., Merkt, F. \& Jungen, C.\ 2004, J. Chem. Phys., 121, 11810

\bibitem{Lee_1973} Lee, C.-M., \& Lu, K.~T.\ 1973, \pra, 8, 1241





\end{thebibliography}

\end{document}